\documentclass[aps,prb,reprint,twocolumn,superscriptaddress,floatfix,longbibliography]{revtex4-2}
\usepackage{preamble}
\usepackage{math_shortcuts}

\begin{document}

\title{Real-frequency quantum field theory applied to the single-impurity Anderson model}

\author{Anxiang Ge}
\thanks{These authors contributed equally to this work.}
\affiliation{\LMU}
\author{Nepomuk Ritz}
\thanks{These authors contributed equally to this work.}
\affiliation{\LMU}
\author{Elias Walter}
\affiliation{\LMU}
\author{Santiago Aguirre}
\affiliation{\LMU}
\author{Jan von Delft}
\affiliation{\LMU}
\author{Fabian B.~Kugler}
\affiliation{\CCQ}
\affiliation{\Rutgers}

\begin{abstract}
A major challenge in the field of correlated electrons is the computation of dynamical correlation functions. For comparisons with experiment, one is interested in their real-frequency dependence. This is difficult to compute, as imaginary-frequency data from the Matsubara formalism require analytic continuation, a numerically ill-posed problem. Here, we apply quantum field theory to the single-impurity Anderson model (AM), using the Keldysh instead of the Matsubara formalism with direct access to the self-energy and dynamical susceptibilities on the real-frequency axis. We present results from the functional renormalization group (fRG) at one-loop level and from solving the self-consistent parquet equations in the parquet approximation. In contrast to previous Keldysh fRG works, we employ a parametrization of the four-point vertex which captures its full dependence on three real-frequency arguments. We compare our results to benchmark data obtained with the numerical renormalization group and to second-order perturbation theory. We find that capturing the full frequency dependence of the four-point vertex significantly improves the fRG results compared to previous implementations, and that solving the parquet equations yields the best agreement with the NRG benchmark data, but is only feasible up to moderate interaction strengths. Our methodical advances pave the way for treating more complicated models in the future. 
\end{abstract}

\maketitle

\section{Introduction}

Strongly correlated electrons are of central interest in condensed matter physics
and a prime application for quantum field theory (QFT).
Two current frontiers in this context are
(i) dealing with two-particle correlations on top of the familiar one-particle correlations,
and (ii) obtaining real-frequency information relevant to experiments, 
as opposed to imaginary-frequency information popular in theoretical analyses.
Indeed, much attention has recently been devoted to the two-particle---or four-point (4p)---vertex of correlated systems, e.g., 
regarding efficient representations 
\cite{Li2016,wentzell_high-frequency_2020,Shinaoka2018,Krien2019,Shinaoka2020,Wallerberger2021,Shinaoka2022}
or the divergences of two-particle irreducible vertices
\cite{Schaefer2013,
Janis2014,
Kozik2015,
Ribic2016,
Schaefer2016c,
Gunnarsson2017,
Vucicevic2018,
Thunstroem2018,
Chalupa2018,
Springer2019,
Melnick2020,
Reitner2020,
Chalupa2021,
Adler2022,
Pelz2023}. 
Moreover, new algorithms have emerged, such as diagrammatic Monte Carlo for real-frequency 2p functions (one frequency argument) working with analytic Matsubara summation 
\cite{PhysRevB.99.035120,
PhysRevB.102.045115,
PhysRevB.104.125114,
PhysRevB.106.035145,
PhysRevLett.129.246401,
Burke2023,
arxiv.2303.04964}
or real-time integration
\cite{PhysRevX.9.041008,
PhysRevB.100.125129,
PhysRevB.103.155104,
nunez_fernandez_learning_2022},
as well as numerical renormalization group (NRG) computations of real-frequency 4p functions (three frequency arguments)
\cite{Kugler2021,Lee2021}. 

Here, we combine aspects (i) and (ii) and study real-frequency two-particle correlations in a QFT framework  within the Keldysh formalism (KF)
\cite{keldysh_diagram_1965,schwinger_brownian_1961,kadanoff_quantum_1962}.
We employ  two related methods: functional renormalization group (fRG) flows at one-loop level \cite{metzner_functional_2012}
and solutions of the self-consistent parquet equations
\cite{bickers_self-consistent_2004}. 
These approaches are promising candidates for real-frequency diagrammatic extensions \cite{Rohringer2018} of dynamical mean-field theory \cite{Georges1996}, where the self-consistently determined impurity model is solved with NRG \cite{Bulla2008}. In practice, this means using the NRG 4p vertex \cite{Kugler2021,Lee2021} as input for 
fRG \cite{taranto_infinite_2014,vilardi_antiferromagnetic_2019}
or the parquet equations \cite{Toschi2007,Held2008}. 
Fully exploiting this non-perturbative input requires taking the full frequency dependence of the 4p vertex into account. 
The present work is a proof-of-principle study showing that it is indeed possible to track 
the three-dimensional \textsl{real-frequency} dependence of the 4p vertex with fRG and parquet methods.

To demonstrate our capability of handling 4p vertices in real frequencies, we choose the well-known \cite{hewson_kondo_1993} single-impurity Anderson model (AM) \cite{anderson_localized_1961} as a test case. Its 4p vertex depends only on frequency and spin arguments, orbital or momentum degrees of freedom are not involved.  Moreover, we can benchmark our results against numerically exact data obtained with NRG \cite{Bulla2008}.

On a historical note, we mention some early pioneering works on the AM 
where multipoint functions depending on multiple real frequencies were computed using various diagrammatic methods \cite{Anders_1994, Anders_1995, Kroha1997, Kroha1998}. Anders and Grewe \cite{Anders_1994, Anders_1995} computed the finite-temperature impurity density of states and spin-fluctuation spectra up to order $\mathcal{O}(1/N^2)$ in a large-$N$ expansion using a resummation that included skeleton diagrams  of the crossing variety up to infinite order. This approach involved the analytic continuation of 2p and 3p functions from imaginary to real frequencies. Kroha, W\"olfle, and Costi \cite{Kroha1997, Kroha1998} studied the AM in the strong-coupling limit using a
slave-boson treatment of local fermions and a conserving $T$-matrix approximation. They computed the auxiliary (pseudofermion and slave boson) spectral functions in the Kondo regime. Their approach involved the analytic continuation of $T$-matrices (4p objects depending on three frequencies) from imaginary to real frequencies. This was possible due to two  simplifications arising in their approach. First, the Bethe--Salpeter equations for the $T$-matrices were simplified via ladder approximations that neglect interchannel feedback but are sufficient to capture the leading and subleading infrared singularities. Second, the auxiliary propagators involve projection factors that cause their contributions to vanish along the branch cuts encountered during the analytic continuation of the $T$-matrices. As a result, only integrations along branch cuts of the conduction electron propagators contribute to the auxiliary-particle self-energies. In particular, only one of the fifteen Keldysh components of the $T$-matrices were involved in these computations. 

In the present paper, we consider a more general setting. We compute the full 4p (impurity-electron) vertex, which requires a treatment of the complete Keldysh structure. Furthermore, the diagrammatic methods considered here---the fRG and the parquet equations---treat all three channels of two-particle fluctuations (density, magnetic, pairing) in an equitable manner, fully including interchannel feedback. The latter causes severe technical complications: each channel has its own frequency parametrization; hence, interchannel feedback involves interpolations between different frequency parametrizations, which in turn demand great care when working with discrete frequency grids.  One of our goals is to develop numerical strategies for conquering these complications in a general, robust manner, as a first step toward studying more complicated models in future work.

Keldysh fRG flows with dynamic 2p and 4p functions were pioneered by Jakobs and collaborators
\cite{jakobs_functional_2010,
jakobs_nonequilibrium_2010,
jakobs_properties_2010}
and subsequently used in Refs.~\onlinecite{schimmel_transport_2017,Schimmel2017,Weidinger2021}.
In all of these works, the dependence of the 4p vertex on three frequencies was approximated by a sum of three functions, each depending on only one (bosonic) frequency.
Here, we present---for the first time---Keldysh one-loop fRG flows 
with the full, three-dimensional frequency dependence of the vertex,
finding remarkable improvement compared to previous implementations 
\cite{jakobs_functional_2010,jakobs_nonequilibrium_2010}.
We also solve the parquet equations in the parquet approximation (PA) in this setting, 
yielding results closest to NRG
in the regime where the parquet self-consistency iteration converges.
This regime corresponds rather accurately to the condition $u \!<\! 1$,
where $u \!=\! U/(\pi\Delta)$ is the dimensionless coupling constant
that controls the (convergent bare) zero-temperature perturbation series \cite{Zlatic1983}.
For completeness, we also discuss second-order perturbation theory (PT2).
Although the PT2 self-energy in the particle-hole symmetric AM (sAM) yields strikingly good results (for known reasons, see Sec.~\ref{sec:benchmark_methods}),
the susceptibilities or the results in the asymmetric AM (aAM) clearly show the benefits of the infinite diagrammatic resummations provided by fRG and the PA.

{A conceptual equivalence between truncated fRG flows and solutions of the parquet equations has been established 
via the multiloop fRG \cite{kugler_multiloop_2018,kugler_multiloop_2018-1,kugler_derivation_2018}. For the AM treated in imaginary frequencies, this equivalence was analyzed numerically in Ref.~\onlinecite{Chalupa2022}, and multiloop convergence was demonstrated up to moderate interaction strengths. We refrain from presenting a similar study in real frequencies here, leaving that for future work.

The rest of the paper is organized as follows: In Sec.~\ref{sec:background}, we give a minimal introduction to the KF (Sec.~\ref{sec:Keldysh})
and summarize the methodical background for fRG and the PA (Secs.~\ref{sec:diagrammatics} and \ref{sec:manybody}). The AM is briefly introduced in Sec.~\ref{sec:AIM},
followed by a concise description of our benchmark methods for this model (Sec.~\ref{sec:benchmark_methods}). 
In Sec.~\ref{sec:results}, we present our results, beginning with dynamical correlation functions computed directly on the real-frequency axis (Sec.~\ref{sec:dynamics}). We then turn to various static properties in Sec.~\ref{sec:static} and check the fulfillment of zero-temperature identities between them (Sec.~\ref{sec:zero-temp-identities}). The frequency-dependent two-particle vertex is shown in Sec.~\ref{sec:vertex}. We conclude in Sec.~\ref{sec:Conclusion} and give a brief outlook on possibilities for future work.

Nine appendices are devoted to technical matters.
Appendix~\ref{app:vertex_parametrization} summarizes our parametrization of the 4p vertex and its symmetry relations. 
Appendix~\ref{app:vertex-components} shows the frequency dependence of all vertex components, as obtained in the PA.
The fully parametrized parquet and fRG flow equations are stated in App.~\ref{app:Fully_Parametrized_Functions} ,
and App.~\ref{app:SDE} discusses a channel-adapted evaluation of the Schwinger--Dyson equation for the self-energy in the PA. Appendix~\ref{app:equal-time} deals with a known equal-time subtlety in the KF, relevant for computing, e.g., the Hartree self-energy in the aAM. In App.~\ref{app:PT2}, we give a concise definition of all diagrammatic contributions to PT2. We provide more details on the actual fRG and PA implementation in App.~\ref{app:implementation} and comment on the numerical costs in App.~\ref{app:numerics}. Finally, App.~\ref{app:chi_m_convergence} scrutinizes the fRG static magnetic susceptibility at $u \gtrsim 1$ for different settings of the frequency resolution.

\section{Background} \label{sec:background}

\subsection{Keldysh formalism}\label{sec:Keldysh}
The Keldysh formalism \cite{keldysh_diagram_1965,schwinger_brownian_1961,kadanoff_quantum_1962}
is a suitable framework for studying systems out of equilibrium,
as well as systems in thermal equilibrium if aiming for a finite-temperature real-frequency description.
An extensive introduction can be found in Ref.~\onlinecite{kamenev_field_2014};
more compact introductions in the context of fRG are also given in related PhD theses \cite{jakobs_functional_2010,schimmel_transport_2017,klockner_functional_2019,walter_keldyshfrg_2022}.
Here, we only give a short summary of the concepts needed in this paper.

Consider a (potentially time-dependent) Hamiltonian $H(t)$ and a density matrix known at time $t_0$, $\rho_0 = \rho(t_0)$. The expectation value of an operator $\hat{O}$ at time $t$ reads
\begin{align}
    \expval{\hat{O}(t)} 
    &= \tr \left[ \tilde{\timeordering} e^{-i \int_{t}^{t_0} \md t' H(t')} \,\hat{O} \, \timeordering e^{-i \int_{t_0}^{t} \md t' H(t')} \, \rho_0 \right] 
    .
\label{eq:ex_value}
\end{align}
Here, $\timeordering$ is the time-ordering operator, and $\tilde{\timeordering}$ denotes anti-time ordering.
In the KF, one rewrites Eq.~\eqref{eq:ex_value} as
\begin{align}
\label{eq:contour}
    \expval{\hat{O}(t)} 
    &= \tr \left[ \timeordering_\contour \left\{ e^{-i \int_{t}^{t_0} \md t' H^+(t')} \,\hat{O} \, e^{-i \int_{t_0}^{t} \md t' H^-(t')} \, \rho_0 \right\} \right] \nonumber \\
    &= 
    \tikzinclude{contour}
    \,.
\end{align}
The Hamiltonian, and all operators in it, acquire an additional contour index $c \!=\! \mp$, indicating whether they sit on the forward ($-$) or backward ($+$) branch of the Keldysh double-time contour. The contour-ordering operator $\timeordering_\contour$ puts all operators on the backward branch left of those on the forward branch, and anti-time orders (time orders) them on the backward (forward) branch.

In the above equation, $\hat{O}$, inserted at time $t$, can be placed on either branch. However, if $\hat{O}$ is a product of multiple operators, they also come with contour indices to ensure the correct ordering. It follows that an $n$-point correlator generically has $2^n$ Keldysh components. For example, the two-point correlator in terms of the creation ($\psi^\dag$) and the annihilation operator ($\psi$) reads 
\begin{align}
G^{c|c'}(t,t') = -i \expval{\timeordering_\contour \, \psi^c(t) \psi^{\dag c'}(t')} \label{eq:G_jj'}
.
\end{align}
Resolving the contour indices $c$, $c'$ yields the matrix
\begin{align}
G^{c|c'} =
\begin{pmatrix}
    G^{-|-} & G^{-|+} \\ G^{+|-} & G^{+|+} 
\end{pmatrix}
=
\begin{pmatrix}
    G^{\timeordering} & G^< \\ G^> & G^{\tilde\timeordering} \label{eq:G_jj'_matrix}
\end{pmatrix}
.
\end{align}

Using the redundancy
$G^< + G^> - G^{\timeordering} - G^{\tilde\timeordering} = 0$, which holds as long as $t\neq t'$ (see App.~\ref{app:equal-time} for the case $t \!=\! t'$), the Keldysh structure of $G$ can be simplified.
The Keldysh rotation invokes the Keldysh indices $k \!=\! 1$ and $2$, where
\begin{align}
    \psi^1 = \tfrac{1}{\sqrt{2}} (\psi^- - \psi^+) \,, \quad
    \psi^2 = \tfrac{1}{\sqrt{2}} (\psi^- + \psi^+) \,,
\end{align}
and equivalently for $\psi^\dag$.
We can thus define a basis transformation matrix $D$ via $\psi^{k} = D^{kc} \psi^c$:
\begin{align}
	D = \tfrac{1}{\sqrt{2}}
	\begin{pmatrix}
		1 & -1 \\ 1 & \phantom{-}1
	\end{pmatrix}
	\,, \quad
	D^{-1} = \tfrac{1}{\sqrt{2}}
	\begin{pmatrix}
		\phantom{-}1 & 1 \\ -1 & 1
	\end{pmatrix}
	\,.
\end{align}
Rotating $G$ as $G^{k|k'} \!=\! D^{kc} G^{c|c'} (D^{-1})^{c'k'}$ yields
\begin{align}
    G^{k|k'}
    =
    \begin{pmatrix}
        G^{1|1} & G^{1|2} \\ G^{2|1} & G^{2|2}
    \end{pmatrix}
    =
    \begin{pmatrix}
        0 & G^A \\ G^R & G^K
    \end{pmatrix}
    \,,
    \label{eq:G_Keldysh_structure}
\end{align}
where $G^{1|1} = 0$ follows from the redundancy mentioned above.
We find the retarded propagator
\begin{align}
G^R(t_1,t_2) = -i\Theta(t_1-t_2) \expval{ \{ c(t_1),c^\dag(t_2) \} }
,
\end{align}
where $\{\cdot,\cdot\}$ denotes the anticommutator,
and its advanced counterpart $G^A(t_1,t_2) \!=\! (G^R(t_2,t_1))^*$,
as well as the Keldysh propagator $G^K(t_1,t_2) \!=\! -(G^K(t_2,t_1))^*$ \cite{jakobs_functional_2010}.

For a time-independent problem, we have $G(t_1,t_2) = G(t_1-t_2)$ and frequency conservation with
\begin{align}
G(\nu) = \int \md t \, e^{i\nu t} \, G(t) \,, \quad
G(t) = \int \frac{\md\nu}{2\pi} \, e^{-i\nu t} \, G(\nu)\,.
\end{align}
In the following, we consider thermal equilibrium at temperature $T$
and chemical potential $\mu$, set to zero.
Then, the density matrix is $\rho_0 \!=\! e^{-H/T} / \partsum$ (with $k_B\!=\!1$ and $\partsum \!=\! \tr e^{-H/T}$), 
and the Keldysh components of $G$ fulfill the fluctuation-dissipation theorem (FDT) \cite{kamenev_field_2014,jakobs_functional_2010}
\begin{align}
    G^K(\nu) 
    &= 2i\tanh \left(\tfrac{\nu}{2T}\right) \Im G^R(\nu) 
    .
    \label{eq:Keldysh_formalism:Field_theory:1P:FDT_GK}
\end{align}

\subsection{Diagrammatic framework}\label{sec:diagrammatics}
The one-particle propagator can be expressed through the bare propagator $G_0$ and the self-energy $\Sigma$ via the Dyson equation.
Using multi-indices $1$, $1'$, etc., we have
\begin{align}
    G_{1|1'} 
    &= 
    \begin{gathered}
    \vspace{2.5ex}
    \tikzinclude{Keldysh_formalism-Dyson_G}
    \end{gathered}
    =
    \begin{gathered}
    \vspace{-1.5ex}
    \tikzinclude{Keldysh_formalism-Dyson_G0}
    \end{gathered}
    +
    \begin{gathered}
    \vspace{-1.5ex}
    \tikzinclude{Keldysh_formalism-Dyson_G0SigmaG}
    \end{gathered}
    ,
    \label{eq:Keldysh_formalism:Field_theory:Dyson_equation}
\end{align}
where the internal arguments $2,$ $2'$ are summed over.
This equation is solved by $G \!=\! (G_0^{-1} \!-\! \Sigma)^{-1}$.
The self-energy has a Keldysh structure similar to Eq.~\eqref{eq:G_Keldysh_structure},
\begin{align}
    \Sigma^{k_1'|k_1}
    =
    \begin{pmatrix}
        \Sigma^{1|1} & \Sigma^{1|2} \\ \Sigma^{2|1} & \Sigma^{2|2}
    \end{pmatrix}
    =
    \begin{pmatrix}
        \Sigma^K & \Sigma^R \\ \Sigma^A & 0
    \end{pmatrix}
    ,
    \label{eq:Keldysh_formalism:Field_theory:1P:Selfenergy}
\end{align}
and
$\Sigma^K(\nu) = 2i\tanh \left(\tfrac{\nu}{2T}\right) \Im \Sigma^R(\nu)$,
cf.\ Eq.~\eqref{eq:Keldysh_formalism:Field_theory:1P:FDT_GK}.

The two-particle (or four-point) correlation function $G^{(4)}$ can be expressed through the four-point vertex $\Gamma$,
\begin{align}
    G^{(4)}_{12|1'2'} 
    &= 
    \tikzinclude{Keldysh_formalism-G4} \nonumber \\
    &=
    \tikzinclude{Keldysh_formalism-G4_dcon1}
    -
    \tikzinclude{Keldysh_formalism-G4_dcon2}
    +
    \tikzinclude{Keldysh_formalism-G4_con}
    \,,
    \label{eq:Keldysh_formalism:Field_theory:4point_function}
\end{align}
where the internal arguments $(3,3',4,4')$ are again summed over.
From $G^{(4)}$, one obtains susceptibilities by contracting pairs of external legs (see App.~\ref{app:Fully_Parametrized_Functions} for details).

The bare vertex, as the full vertex, is fully antisymmetric in its indices. Thus, a purely local and instantaneous interaction is of the type 
\begin{align}
    (\Gamma_0)_{\sigma_1'\sigma_2'|\sigma_1\sigma_2} & (t_1',t_2'|t_1,t_2) 
    = - U \delta(t_1'\!=\!t_2'\!=\!t_1\!=\!t_2) \delta_{\sigma_1,\bar\sigma_2'} 
    \nonumber \\
    & \, \times (\delta_{\sigma_1',\sigma_2}\delta_{\sigma_2',\sigma_1} - \delta_{\sigma_1',\sigma_1}\delta_{\sigma_2',\sigma_2}) 
    ,
\end{align}
with $U>0$ for a repulsive interaction.
This corresponds to a Hugenholtz diagram (single dot) \cite{negele_quantum_1998}
\begin{align}
    (\Gamma_0)_{1'2'|12} &= 
    \tikzinclude{Keldysh_formalism-Gamma0}
    =
    \tikzinclude{Keldysh_formalism-Gamma0_e}
	-
    \tikzinclude{Keldysh_formalism-Gamma0_d}
    \!.
    \label{eq:Keldysh_formalism:Field_theory:G0_Gamma0}
\end{align}

As the bare vertex is part of either $H^+$ or $H^-$ in Eq.~\eqref{eq:contour}, all its 
contour indices must be equal \cite{jakobs_functional_2010},
\begin{align}
    (\Gamma_0)_{1'2'|12} 
    &=
    -c_1 \, \delta_{c_1'=c_2'=c_1=c_2}
    (\Gamma_0)_{\sigma_1'\sigma_2'|\sigma_1\sigma_2}(t_1',t_2'|t_1,t_2) 
    .
\end{align}
It acquires a minus sign $-c_1$ when moved from the forward ($c_1=-$) to the backward ($c_1=+$) branch of the Keldysh contour.
After Keldysh rotation, one obtains 
\begin{align}
    (\Gamma_0)^{k_1'k_2'|k_1k_2}_{\sigma_1'\sigma_2'|\sigma_1\sigma_2} 
    = 
    \left\lbrace
    \begin{array}{ll}
        \tfrac12 (\Gamma_0)_{\sigma_1'\sigma_2'|\sigma_1\sigma_2} , \ & \sum_i k_i \text{  odd} ,
        \\
        0, \ & \text{else}, 
    \end{array}
    \right.
    \label{eq:Keldysh_formalism:Field_theory:2P:Bare_vertex_Keldysh_indices}
\end{align}
where $\sum_i k_i$ is short for $k_1'+k_2'+k_1+k_2$.

\subsection{Many-body approaches}\label{sec:manybody}
So far, we defined the basic objects of interest, namely one- and two-particle correlation functions in the KF, 
encapsulated in the self-energy $\Sigma$ and the 4p vertex $\Gamma$,
\begin{flalign}
    \Sigma_{1'|1}
    & =
    \tikzinclude{Keldysh_formalism-Sigma}
    \,,\quad
    \Gamma_{1'2'|12}
    =
    \tikzinclude{Keldysh_formalism-Gamma}
    \,.
    \hspace{-0.5cm} &
\end{flalign}
One can derive a diagrammatic perturbation series for each of them.
However, to extend our description from weak to intermediate coupling, we want to resum infinitely many diagrams.
We use two strategies achieving this: 
fRG \cite{metzner_functional_2012,kopietz_introduction_2010}
and the PA \cite{bickers_self-consistent_2004}. 
We summarize both schemes in turn and then comment on their relation.

In fRG, one introduces a scale parameter $\Lambda$ into the bare propagator $G_0$,
such that the theory is solvable at an initial value $\Lambda \!=\! \Lambda_i$,
while the original problem is recovered at a final value $\Lambda \!=\! \Lambda_f$ (i.e., $G_0^{\Lambda_f} \!=\! G_0$).
Here, we choose $G_0^{\Lambda_i}$ very small,
so that $\Sigma^{\Lambda_i}$ and $\Gamma^{\Lambda_i}$
can be obtained by perturbation theory
or by iterating the parquet equations (see below) until convergence.
The final results $\Sigma^{\Lambda_f} \!=\! \Sigma$ and $\Gamma^{\Lambda_f} \!=\! \Gamma$ are obtained by solving a set of flow equations.
In fact, the fRG provides an infinite hierarchy of flow equations, which is in principle exact but must be truncated in practice.
The flow equations for $\dot{\Sigma} \!=\! \partial_\Lambda \Sigma$ and $\dot{\Gamma} \!=\! \partial_\Lambda \Gamma$ in diagrammatic notation are
\begin{subequations}
\label{eq:mfRG:fRG:Flow_equations}
\begin{align}
    \dot{\Sigma} \
    &= \
    -
    \tikzinclude{mfRG-1l-dSigma}
    \,,
    \label{eq:mfRG:fRG:Flow_equations:Selfenergy_flow}
    \\
    \dot{\Gamma} \
    &= \
    \tikzinclude{mfRG-1l-dGamma_a}
    + \frac{1}{2} \!\!
    \tikzinclude{mfRG-1l-dGamma_p}
    \nonumber \\ 
    &- \
    \tikzinclude{mfRG-1l-dGamma_t}
    \ + \
    \tikzinclude{mfRG-1l-dGamma_6}
    \ .
    \label{eq:mfRG:fRG:Flow_equations:Vertex_flow}
\end{align}
\end{subequations}
The propagator with a dash is the single-scale propagator
$S = \partial_\Lambda G |_{\Sigma=\text{const.}}$;
propagator pairs with a dash indicate
$\dot{\Pi}^S = S G + G S$.
We adopt the one-loop fRG scheme where the truncation consists of $\Gamma^{(6)} \!\approx\! 0$.
As is commonly done, we then employ the so-called Katanin substitution \cite{katanin_fulfillment_2004}
where $\dot{\Pi}^S$ is replaced by $\dot{\Pi} = \dot{G} G + G \dot{G}$.

The parquet formalism consists of solving a self-consistent set of equations on the one- and two-particle level. 
It involves the Schwinger--Dyson equation (SDE)
\begin{subequations}
\label{eq:all_parquet_equations}
\begin{align}
    \begin{gathered}
    \vspace{-1ex}
    \tikzinclude{mfRG-parquet-SDE}
    \end{gathered}
    =
    -
    \tikzinclude{mfRG-parquet-SDE0}
    -
    \frac{1}{2} \
    \tikzinclude{mfRG-parquet-SDE_Gamma}
    ,
    \label{eq:mfRG:mfRG:Parquet:SDE}
\end{align}
where the first term is the Hartree self-energy $\Sigma_{\mathrm{H}}$,
as well as the Bethe--Salpeter equations (BSEs)
\begin{align}
\label{eq:parquet:BSEa}
    \tikzinclude{mfRG-parquet-BSE_a}
    &=
    \phantom{\frac{1}{2} \!\!}
    \quad
    \tikzinclude{mfRG-parquet-BSE_a1}
    ,
    \\
\label{eq:parquet:BSEp}
    \tikzinclude{mfRG-parquet-BSE_p}
    &=
    \frac{1}{2} \!\!
    \tikzinclude{mfRG-parquet-BSE_p1}
    ,
    \\
\label{eq:parquet:BSEt}
    \tikzinclude{mfRG-parquet-BSE_t}
    &=
      \ - \ \
    \tikzinclude{mfRG-parquet-BSE_1} \, 
    .
\end{align}
Here, $\gamma_r$ is the two-particle reducible vertex in a given channel $r \!\in\! \{a, p, t\}$,
while $I_r \!=\! \Gamma - \gamma_r$ is the corresponding two-particle irreducible vertex. The parquet equation
\begin{align}
\Gamma = R + \gamma_a + \gamma_p + \gamma_t
\label{eq:parquet_equation}
\end{align}
\end{subequations}
gives the full vertex in terms of the two-particle reducible vertices as well as the fully irreducible vertex $R$.
The set of equations~\eqref{eq:all_parquet_equations} is exact. 
However, $R$ in Eq.~\eqref{eq:parquet_equation} is not determined by an integral equation itself and serves as an input, for which an approximation must be used in practice.
The PA is the simplest such approximation:
\begin{align}
R = \Gamma_0 + \mathcal{O}[(\Gamma_0)^4] \approx \Gamma_0
.
\end{align}
Thus, the set of equations \eqref{eq:all_parquet_equations} closes and can be solved by standard means.

The truncated (one-loop) fRG flow and the PA are closely related but differ in details. An equivalence between them is established by the multiloop fRG \cite{kugler_multiloop_2018,kugler_multiloop_2018-1,kugler_derivation_2018}
(see also Refs.~\onlinecite{tagliavini_multiloop_2019,hille_quantitative_frg_2020,thoenniss_multiloop_2020,Kiese2022,Chalupa2022,gievers_multiloop_2022,Ritter2022}):
By incorporating additional terms into the flow equations, which simulate part of the intractable six-point vertex in the fRG hierarchy of flow equations, the scale derivative of the self-energy and vertex is completed to a total derivative of diagrams, which are precisely the diagrams contained in the PA. Hence, if multiloop convergence can be achieved, the regulator dependence of the truncated fRG flow is eliminated, and one obtains results equivalent to the PA. Here, we restrict ourselves to one-loop fRG flows. Our numerical explorations with multiloop fRG for the AM in the KF have so far shown that the additional terms are numerically less well behaved, requiring a prohibitively high numerical resolution. This task is therefore left for future work, where compression techniques such as the new quantics tensor cross interpolation scheme \cite{nunez_fernandez_learning_2022, Shinaoka2023, ritter_quantics_2023} could be used to keep the needed numerical resources manageable.

\subsection{Single-impurity Anderson model}\label{sec:AIM}
The formalism introduced above is completely general
and can be applied, e.g., to lattice or impurity models alike.
Comparing Keldysh to Matsubara approaches, 
the spatial or momentum degrees of freedom of lattice models are treated similarly in both cases.
By contrast, the temporal or frequency dynamics are naturally very different.
In impurity models, the frequency dynamics are isolated,
saving the cost of including momentum variables. 
Hence, we consider in this paper the AM \cite{anderson_localized_1961} in thermal equilibrium.
Its physical behavior is well understood \cite{hewson_kondo_1993},
and NRG \cite{Bulla2008}
can be used to obtain highly accurate real-frequency benchmark data.

The model is defined by the Hamiltonian
\begin{align}
H &= 
\sum_{\epsilon\sigma} \epsilon c_{\epsilon\sigma}^\dagger c_{\epsilon\sigma}
+
 (\epsilon_d + h) n_\uparrow + (\epsilon_d - h) n_\downarrow
+ U n_\uparrow n_\downarrow 
\nonumber \\
&\quad +
\sum_{\epsilon\sigma} (V_\epsilon d_\sigma^\dagger c_{\epsilon\sigma} + \mathrm{H.c.})
,
\label{eq:SIAM_Hamiltonian}
\end{align}
with spinful bath electrons, created by $c_{\epsilon\sigma}^\dagger$, and a local level ($d_\sigma^\dagger$). The latter has an on-site energy $\epsilon_d$ and Coulomb repulsion $U$ acting on $n_\sigma \!=\! d_\sigma^\dag d_\sigma$. Although we consider $h \!=\! 0$, we include the magnetic field in Eq.~\eqref{eq:SIAM_Hamiltonian} for a simple definition of the magnetic susceptibility.
The bath electrons are integrated out, yielding the frequency-dependent retarded hybridization function
$-\Im \Delta^R(\nu) = \sum_\epsilon \pi |V_\epsilon|^2 \delta(\nu-\epsilon)$.
We consider a flat hybridization in the wide-band limit, $\Delta^R_\nu \!=\! -i\Delta$,
so that the bare impurity propagator reads
\begin{align}
G_{0}^{R}(\nu) 
&= 
\frac{1}{\nu - \epsilon_d + i\Delta}
.
\end{align}
We use the dimensionless parameter $u \!=\! U/(\pi\Delta)$ for the interaction strength on the impurity in units of the hybridization strength to the bath. We focus on two choices of
the on-site energy: one with particle-hole symmetry, $\epsilon_d \!=\! -U/2$, and one without, $\epsilon_d \!=\! 0$. We refer to these as the
symmetric AM (sAM) and asymmetric AM (aAM), respectively.

For the sAM, $\Sigma_\mH \!=\! U/2$ is conveniently absorbed into the bare propagator,
\begin{align}
G^R_0 \to G^R_\mH = \frac{1}{\nu - \epsilon_d + i\Delta - \Sigma_\mH}
= \frac{1}{\nu + i\Delta}
.\label{eq:Hartree-propagator}
\end{align}
For perturbative calculations in the aAM
(as in PT2 or to initialize the parquet iterations), 
we also replace $G_0$ by $G_\mH$
(see App.~\ref{app:equal-time} for details).

For the fRG treatment, we use the hybridization flow \cite{jakobs_functional_2010},
where $\Delta$ acts as the flow parameter and is decreased from a very large value to the actual value of interest. This is convenient because every point of the flow describes a physical system, at the given values of $\Delta$, $U$, $T$. In other words, the fRG flow provides a complete parameter sweep in $\Delta$, while the other parameters ($U$, $T$) remain fixed.
Then, the fRG single-scale propagator is
\begin{align}
S^{R}(\nu) 
&= \partial_\Delta G^{R}(\nu) \big|_{\Sigma=\text{const.}} 
= - i [G^{R}(\nu)]^2 
. \label{eq:single-scale-hyb}
\end{align}
In the limit $\Delta \to \infty$, the values of $\Gamma$ and $\Sigma$ are \cite{jakobs_functional_2010}
\begin{align}
\Gamma \big|_{\Delta=\infty} &= \Gamma_0 
, \qquad
\Sigma^{R} \big|_{\Delta=\infty} = \Sigma_\mH = U \langle n_{\sigma} \rangle
.
\end{align}
Note that while all vertex diagrams of second order or higher vanish as $\Delta \to \infty$, the first-order contribution of $\Sigma^{R/A}$ (the Hartree term $\Sigma_\mH$) is finite. As discussed in App.~\ref{app:equal-time}, $\Sigma_\mH$ is given by an integral over $G^<$, which gives a finite value $U\langle n_{\sigma} \rangle$ even in the limit $\Delta\rightarrow\infty$. In practice, we start the flow at a large but finite value of $\Delta$, and use the self-consistent solution of the parquet equations as the initial conditions for $\Sigma$ and $\Gamma$, as they can be easily obtained for sufficiently large $\Delta$.

\subsection{Benchmark methods}
\label{sec:benchmark_methods}
As a real-frequency benchmark method,
we use NRG in a state-of-the-art implementation
based on the \mbox{QSpace} tensor library
\cite{weichselbaum_non-abelian_2012,weichselbaum_tensor_2012,Weichselbaum2020}.
We employ a discretization parameter of $\Lambda \!=\! 2$, 
average over $n_z \!=\! 6$ shifts of the logarithmic discretization grid \cite{Zitko2009}, 
and keep up to 5000 SU(2) multiplets during the iterative diagonalization.
Dynamical correlators are obtained via the full density-matrix NRG 
\cite{weichselbaum_sum-rule_2007,Peters2006}, 
using adaptive broadening \cite{lee_adaptive_2016,Lee2017}
and a symmetric improved estimator for the self-energy \cite{Kugler2022}.
We also extract zero-temperature quasiparticle parameters from the NRG low-energy spectrum
\cite{Hewson2004,Bauer2007,Nishikawa2010,Nishikawa2010-1,Oguri2011,Kugler2020,Tsutsumi2023}.
Dividing the quasiparticle interaction $\tilde{U}$ by the square of the quasiparticle weight $Z^2$ yields the 4p vertex at vanishing frequencies $\Gamma_{\uparrow\downarrow}(\vec{0})$.
Thereby, we obtain $\Gamma_{\uparrow\downarrow}(\vec{0}) = -\tilde{U}/Z^2$ at $T \!=\! 0$ very efficiently and accurately.
For a finite-temperature estimate, we divide $\tilde{U}$ by the finite-temperature $Z$ deduced from the dynamic self-energy as opposed to the zero-temperature $Z$ following from the low-energy spectrum.
We also compute the dynamical 4p vertex in the Keldysh formalism, exploiting the recent advances described in Refs.~\onlinecite{Kugler2021,Lee2021}.

For completeness, we also compare our results to PT2.
Pertubation theory of the AM is known to work well when expanding around the nonmagnetic Hartree--Fock solution
\cite{yosida_perturbation_1970,yamada_perturbation_1975,yosida_perturbation_1975,yamada_perturbation_1975-1,Zlatic1983}.
PT2 famously and fortuitously 
(cf.~the iterated perturbation theory in the DMFT context \cite{Georges1996})
gives \textsl{very} good results for the self-energy of the sAM, where $\epsilon_d \!=\! -U/2$ and $\Sigma_{\mathrm{H}}$ cancel exactly.
The reason is that $\Sigma_{\mathrm{PT}2}$ is correct in the limits $u \!\to\! 0$
\textsl{and} $u \!\to\! \infty$.
In the latter case, the spectrum $-\tfrac{1}{\pi}\mathrm{Im}G^R$ consists of two discrete peaks,
and, in the sAM, the resulting expression for $\Sigma^R \!=\! 1/G_0^R - 1/G^R$ is $(U/2)^2/(\nu \!+\! \mi 0^+)$, coinciding with PT2.
One may further note that corrections to $\Sigma_{\mathrm{PT}2}$ start at order $u^4$,
as only even powers contribute to the expansion of $\Sigma$ for the sAM,
and that the expansion converges very quickly (see Fig.~3.6 and 3.7 in Ref.~\onlinecite{yamada_perturbation_1975}).
Additionally, the high-frequency asymptote $\lim_{\nu\to\infty} \nu (\Sigma^R \!-\! \Sigma_{\mathrm{H}})$ 
is fully captured by PT2,
as the general expression $U^2 \langle n_{\sigma} \rangle (1-\langle n_{\sigma} \rangle)$ 
reduces to $(U/2)^2$ (with $\langle n_{\sigma} \rangle \!=\! 1/2$ in the sAM), i.e.,
the second-order result.

For the aAM, $\Sigma_{\mathrm{H}}$ must first be determined in a self-consistent way. 
This is crucial as $\langle n_\sigma \rangle$ is not well approximated by few orders in $u$
(recall the Friedel sum rule at $T\!=\! 0$ \cite{Langreth1966},
$\langle n_\sigma \rangle = \tfrac{1}{2} \!-\! \tfrac{1}{\pi} \arctan [(\epsilon_d+\Sigma(0))/\Delta]$).
The self-consistent Hartree propagator fulfills the Friedel sum rule  at $T \!=\! 0$,
but the resulting $\langle n_\sigma \rangle$ for given $\epsilon_d$ is of course not exact.
When using PT2, we compute quantities of interest, such as $\Sigma_{\mathrm{PT2}}$, using the Hartree propagator
(see App.~\ref{app:PT2} for details). 
However, in contrast to the sAM,
$\Sigma_{\mathrm{PT2}}$ is not exact at $u \!\to\! \infty$
(cf.~Ref.~\onlinecite{Kajueter1996}),
odd powers in $u$ contribute to $\Sigma$, 
and the high-frequency asymptote of $\Sigma_{\mathrm{PT}2}$,
involving $\langle n_\sigma \rangle$, 
is not reproduced exactly.

Finally, we also compare our fRG and PA results to ``K1SF calculations'' mimicking the previous state of the art in Keldysh fRG.
References~\onlinecite{jakobs_nonequilibrium_2010,jakobs_functional_2010,Schimmel2017} 
used a scheme where the full vertex is decomposed into the three channels
[cf.~Eq.~\eqref{eq:parquet_equation}]
and, for each two-particle reducible vertex $\gamma_r$, 
only the dependence on the bosonic transfer frequency is retained
(see Eq.~(76) in Ref.~\onlinecite{jakobs_nonequilibrium_2010}):
\begin{align}
    \label{eq:K1SF_approximation:vertex}
    \Gamma
    \approx
    \Gamma_0 + \sum_{r=a,p,t} \gamma_r(\omega_r)
    .
\end{align}
Note that, within Matsubara fRG, Ref.~\onlinecite{karrasch_SIAM_2008} compared this simplification (called ``Appr.~1'' therein) to the full parametrization.
When inserting the vertex parametrized according to Eq.~\eqref{eq:K1SF_approximation:vertex} into the self-energy flow~\eqref{eq:mfRG:fRG:Flow_equations:Selfenergy_flow},
no further approximations are needed.
However, when inserting the vertex on the right of the vertex flow equation~\eqref{eq:mfRG:fRG:Flow_equations:Vertex_flow}, the inter-channel contributions are approximated by their static values
(in thermal equilibrium with $\mu = 0$, see Eq. (83) in Ref.~\onlinecite{jakobs_nonequilibrium_2010})
\begin{align}
    \Gamma \Big|_{\mathrm{RHS}(\gamma_r)}
    \approx
    \Gamma_0 + \gamma_r(\omega_r) +
    \sum_{r' \neq r} \gamma_{r'}(\omega_{r'})\Big|_{\omega_{r'}=0}
    .
\end{align}
With this approximation the only frequency dependence of the integrands lies in the propagator pair.
By contrast, an exact decomposition of each $\gamma_r$ has the form \cite{wentzell_high-frequency_2020}
\begin{align}
\label{eq:asymptotic_decomposition}
\gamma_{r}(\omega_r,\nu_r,\nu_r') & = K_{1r}(\omega_r) + K_{2r}(\omega_r,\nu_r) 
\nonumber \\
& \ 
+ K_{2'r}(\omega_r,\nu'_r) + K_{3r}(\omega_r,\nu_r,\nu_r').    
\end{align}
(The frequency arguments $\omega_r$, $\nu_r$, $\nu'_r$ are defined in App.~\ref{app:vertex_parametrization}, Fig.~\ref{fig:parametrization:frequency_convention}.)
Thus, the above approximation can be understood by retaining only the $K_{1r}$ contributions
while ensuring a static feedback (SF) between the different channels---hence the abbreviation \SF.

\begin{figure*}[ht]
\centering
\begin{minipage}[t]{.482\textwidth}
\centering
\includegraphics[width=1.02\textwidth]{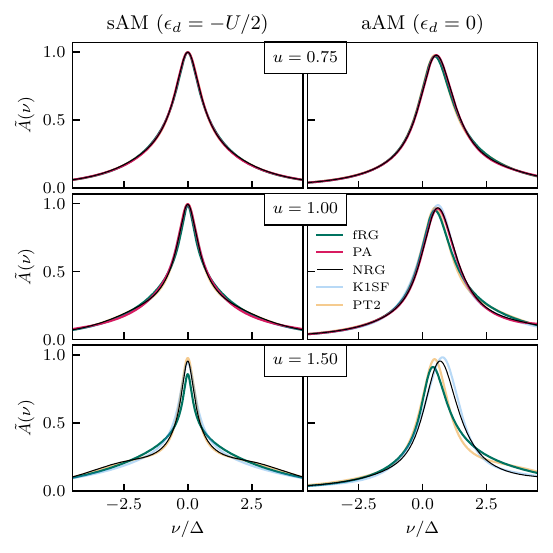}
    \caption{Spectral functions $\tilde{A}(\nu) \equiv \pi\Delta A(\nu)$ for three interaction values $u$ in the symmetric AM (sAM, left) and the 
    asymmetric AM (aAM, right). Deviations between the methods appear with increasing $u$. Here and in all subsequent figures,
we consider a temperature of fixed $T/U=0.01$. At $u \!=\! 1.5$ in the sAM, the onset of Hubbard bands centered at $\nu=\pm U/2$ is only captured by NRG and (for reasons explained in Sec.~\ref{sec:benchmark_methods}) PT2. At this interaction strength, fRG underestimates the quasiparticle peak, and we were unable to converge the PA results.}
    \label{fig:spectral}
\end{minipage}\hfill
\begin{minipage}[t]{.482\textwidth}
\centering
\includegraphics[width=1.02\textwidth]{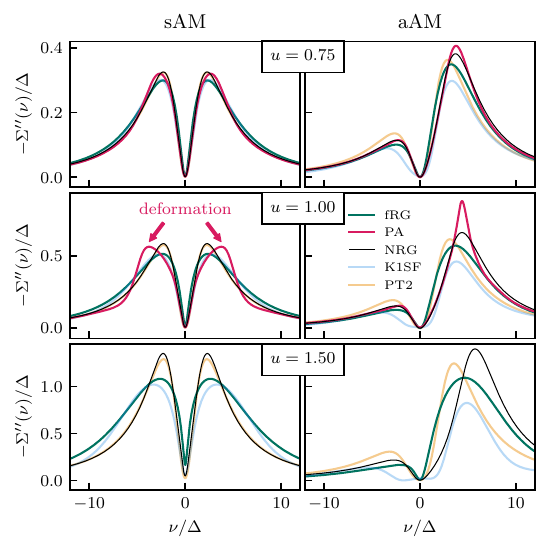}
        \caption{Imaginary part of the retarded self-energy, organized as in Fig.~\ref{fig:spectral}.
        The limitations of PT2 in the aAM are clearly exposed. The PA results are closest to NRG at $u \!=\! 0.75$ for both sAM and aAM, and at $u \!=\! 1$ for the aAM (this corresponds to the regime of not too strong correlation, $Z \gtrsim 0.8$, see Fig.~\ref{fig:QP}).  Artifacts appear at $u \!=\! 1$ in the sAM (where $Z \!\approx\! 0.65$, see Fig.~\ref{fig:QP}). Throughout, the fRG results with full frequency dependence match NRG better than those in the K1SF simplification.
        }
        \label{fig:self-energy}
\end{minipage}
\end{figure*}

Within \SF, there are different ways of treating the feedback from the self-energy. 
Previous works found better results at $T \!=\! 0$ by inserting only the static rather than full dynamic $\Sigma$ into the propagator \cite{jakobs_properties_2010}. We confirm this finding at $T=0$ but observed that the static $\Sigma$ feedback has problems at $T \!\neq\! 0$, failing, e.g., the requirement $\mathrm{Im}\,\Sigma \!<\! 0$. Instead, we obtained much better results (particularly fulfilling $\mathrm{Im}\,\Sigma \!<\! 0$) by using the full dynamic $\Sigma$ feedback together with the Katanin substitution \cite{katanin_fulfillment_2004}.

\subsection{Note on the numerical implementation}
Compared to the more common Matsubara formalism (MF), the KF entails notable differences in the numerical implementation that we summarize here (see App.~\ref{app:implementation} for details).
Most importantly, while finite-temperature Matsubara computations employ a discrete set of (imaginary) frequencies, Keldysh functions depend on continuous (real) frequencies. 
Furthermore, the Keldysh index structure increases the number of components of the correlators (to be computed and stored) by a factor of 4 and 16 for 2p and 4p objects, respectively.
Hence, working in the KF requires considerably higher effort in terms of implementational complexity and numerical resources.

To minimize systematic numerical errors, a faithful representation of the vertex functions is essential.
The decomposition~\eqref{eq:asymptotic_decomposition} of the reducible vertices \cite{wentzell_high-frequency_2020} is beneficial for capturing the high-frequency asymptotics. Indeed, the lower-dimensional asymptotic functions, $K_1$ and $K_{2^{(\prime)}}$, allow for a good resolution at comparably low numerical cost.
A good resolution of the continuous Keldysh functions further necessitates a suitable choice of sampling points.
We use a frequency grid with high resolution at small frequencies, where the vertices show sharp features, and fewer points at higher frequencies.
In fRG with the hybridization flow, the frequency grids also have to be rescaled to account for changes scaling with $\Delta$; for fully adaptive grids (which were not required in this work, cf.~App.~\ref{app:implementation}) see also Refs.~\onlinecite{thoenniss_multiloop_2020,Ritter2022,Kiese2022}. 

Continuous-frequency computations also require efficient integration routines.
We use an adaptive quadrature routine to capture the essential features of sharply peaked functions (cf.~App.~\ref{app:implementation}).
The additional numerical costs due to the Keldysh index structure can be mitigated by vectorization, i.e., by exploiting the matrix structure of the summation over Keldysh components.
Storing all Keldysh components contiguously in memory allows for efficient access to matrix-valued vertex data, which can be combined to matrix-valued integrands via linear algebra operations. (Note that vectorization over Keldysh components requires a quadrature routine that accepts matrix-valued integrands.)
Symmetries are used to reduce the data points that are computed directly, and most resulting symmetry relations are compatible with vectorization over Keldysh indices (see App.~\ref{app:vertex_parametrization}) 

Lastly, the fRG and the parquet solver generally have the advantage that computations can be parallelized efficiently over all combinations of external arguments. We use  OMP and MPI libraries to parallelize execution across multiple CPUs and compute nodes.

\section{Results}\label{sec:results}
In the results, we focus on retarded correlation functions like $G^R$, $\Sigma^R$, and $\chi^R$. For brevity, we denote the real and imaginary parts of, say, $G^R$ by $G'$ and $G''$, respectively, i.e., $G^R \!=\! G' \!+\! i G''$.
Since the fRG flow varies $\Delta$ at fixed $U$ and $T$, 
we consider a temperature of $T/U \!=\! 0.01$. 
Most plots show results both for
the sAM ($\epsilon_d = -U/2$) and aAM ($\epsilon_d \!=\! 0$). Recall that $u \!=\! U/(\pi\Delta)$. 

\subsection{Dynamical correlation functions}\label{sec:dynamics}
As a first quantity that is directly measurable in experiment, we show in Fig.~\ref{fig:spectral} the spectral function $\tilde{A}(\nu) \!\equiv\! \pi\Delta A(\nu) \!=\! -\Delta G''(\nu)$. The absorbed factor of $\pi\Delta$ ensures $\tilde{A}(0) \!=\! 1$ for the sAM and $T \!\to\! 0$. We consider three values of $u\in\{0.75, 1, 1.5\}$, referred to as ``small'', ``intermediate'', and ``large'' in the following
(although truly large interactions in the AM rather are $u\!\gtrsim\! 2$ \cite{Zlatic1983}). There are no PA results for large $u$, as we could not converge the real-frequency self-consistent parquet solver there. 

At small $u$, the curves produced by all methods are almost indistinguishable. 
Small but noticeable deviations occur for the aAM at intermediate $u$, and pronounced deviations are found at large $u$. 
At $u \!=\! 1.5$ in the sAM,
only the methods exact in the $u \!\to\! \infty$ limit 
(cf.\ Sec.~\ref{sec:benchmark_methods}), NRG and PT2, 
produce notable Hubbard bands centered at $\nu \!=\! \pm U/2$,
while fRG also underestimates the height of the quasiparticle peak. 
Nevertheless, one may come to the conclusion that all methods agree to a reasonable degree of accuracy. Note, though, that at small $u$, where $\Sigma$ is small, $G^{R} = 1/([G_0^{R}]^{-1} \!-\! \Sigma^{R})$ and thus also $A(\nu)$ are dominated by the bare propagator.
As all nontrivial features of $A(\nu)=\frac{1}{\pi}\frac{\Delta-\Sigma''(\nu)}{[\nu-\epsilon_d-\Sigma'(\nu)]^2+[\Delta-\Sigma''(\nu)]^2}$ come from $
\Sigma$, we can gain more insight by looking at $\Sigma$ directly.

In Fig.~\ref{fig:self-energy}, we plot the negative imaginary part of the retarded self-energy $-\Sigma''(\nu)$ in units of the hybridization strength $\Delta$. This quantity is strictly non-negative \cite{Kugler2022}, which is a useful and non-trivial consistency check for all our methods. Here, deviations between the methods are visible at each value of $u$. At small $u$, the results mostly agree, albeit better for the sAM than for the aAM. 
At small and intermediate $u$ in the aAM, the PA matches NRG most closely and also captures the peak position correctly, in contrast to fRG, K1SF, and PT2. Strikingly, though, for intermediate $u$ in the sAM (which is the more strongly correlated setting with lower quasiparticle weight $Z$, see Fig.~\ref{fig:QP}), the PA shows considerable deviations from NRG:
$\Sigma''$ has a ``deformation'' in that its maxima are misplaced outward. 
We performed a separate PA calculation in the MF to confirm that the corresponding MF result perfectly matches the ``trivial'' analytic continuation from KF to MF, $-\frac{1}{\pi}\int \mathrm{d}\nu' \frac{\Sigma''(\nu')}{i\nu-\nu'}$, see Fig.~\ref{fig:AC}. Hence, we conclude that the Keldysh self-energy did not acquire artifacts during the real-frequency self-consistent parquet iteration. Instead, the deformations are a deficiency of the PA solution at $u\!=\! 1$, which are obvious in our Keldysh results, but could not have been guessed from the more benign Matsubara self-energy (Fig.~\ref{fig:AC}).

We also observe from Fig.~\ref{fig:self-energy} that
the PT2 results become much worse as soon as one leaves the special case of particle-hole symmetry (see Sec.~\ref{sec:benchmark_methods}). 
The results from fRG with full frequency dependence are better than the ones from K1SF, showing that the frequency dependence of $\Sigma$ is only generated correctly if the dependence of the 4p vertex on its three frequencies is kept \cite{metzner_functional_2012}. In fact, for large $u$ in the aAM, the K1SF result becomes \textsl{negative} (with values on the order of $10^{-5}$) at around $\nu/\Delta\simeq \pm 2$, thus failing the previously mentioned consistency check.

The inadequacies of a constant vertex manifest themselves even in the constant Hartree part of the self-energy, $\Sigma_{\mathrm{H}} \!=\! U \langle n_\sigma \rangle$, shown in Fig.~\ref{fig:hartree}. The fRG and PA calculations produce the NRG value almost exactly, but the \SF curve starts to deviate early. We attribute this to the fact that diagrammatic contributions beyond the $K_1$ level are neglected, introducing an error of $\mathcal{O}(U^3)$ into the flow of $\Sigma$, including $\Sigma_{\mathrm{H}}$, see Eq.~\eqref{eq:hartree-flow}. The PT2 curve shows the converged values obtained from self-consistent evaluations of the Hartree diagram (see App.~\ref{app:equal-time}), which enters the Hartree propagator used in all PT2 computations. The self-consistency is likely the reason why PT2 performs better than \SF (which does not obey such a self-consistency) for small and intermediate $u$. 

Apart from $\tilde{A}$ and $\Sigma$, other dynamical quantities of interest are susceptibilities. In the diagrammatic methods, these are derived directly from the 4p vertex (see App.~\ref{app:Fully_Parametrized_Functions}). We consider the imaginary part of the retarded magnetic and density dynamical susceptibilities $\tilde{\chi}_{\mathrm{m}/\mathrm{d}}(\omega) \!\equiv\! \pi\Delta\chi_{\mathrm{m}/\mathrm{d}}(\omega)$, paying special attention to the peak position and height. The peak position of $\tilde{\chi}_{\mathrm{m}}$ shown in Fig.~\ref{fig:chi_sp} is proportional to the Kondo temperature and decreases with increasing $u$ in the sAM. 
All methods apart from \SF produce good results at small $u$ with only minor deviations from NRG. The deviations are smallest in PA from small to intermediate $u$, until the PA results are no longer available at large $u$. fRG produces reasonable curves but, at large $u$, under- or overestimates the peak in the sAM and aAM, respectively. \SF does not produce sensible results for any $u$ considered, while PT2 performs well for the aAM but yields worse results than fRG in the sAM.

\begin{figure}[t]
    \centering
\includegraphics[width=0.5\textwidth]{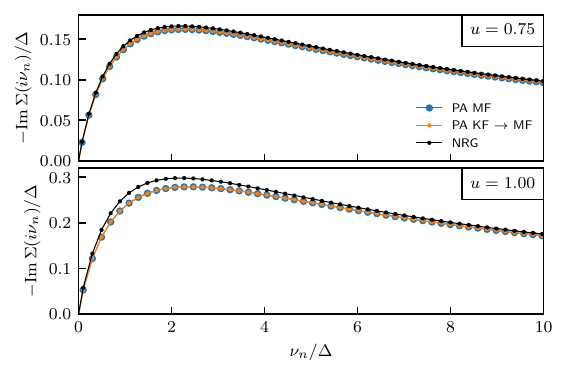}
    \caption{Imaginary part of the Matsubara self-energy in NRG and the PA. The PA results stem from an independent solver implemented in the MF and from the ``trivial'' analytic continuation of $\Sigma''$ obtained in the KF. The \textsl{qualitative} difference between NRG and PA observed in the real-frequency results of Fig.~\ref{fig:self-energy} at $u=1$ can hardly be guessed from these imaginary-frequency results.}
    \label{fig:AC}
\end{figure}

\begin{figure}[t]
    \centering
\includegraphics[width=0.5\textwidth]{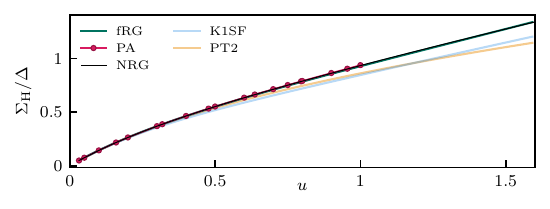}
    \caption{Hartree self-energy $\Sigma_{\mathrm{H}} = U \langle n_\sigma \rangle$ in the aAM. PT2 corresponds to self-consistent solutions of the Hartree term. Only fRG and PA agree well with NRG.}
    \label{fig:hartree}
\end{figure}

\begin{figure*}[ht]
\centering
\begin{minipage}[t]{.482\textwidth}
\centering
\includegraphics[width=1.02\textwidth]{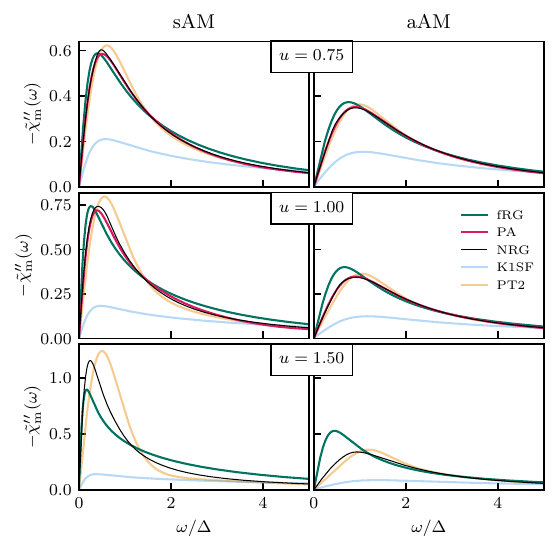}
    \caption{Imaginary part of the dynamical magnetic susceptibility, $\tilde{\chi}_{\mathrm{m}}(\omega) \!\equiv\! \pi\Delta \chi_\mathrm{m}(\omega)$. At small to intermediate $u$, all methods (except \SF) produce good results, while PA matches NRG best. Toward large $u$, fRG does not capture the peak correctly. PT2 performs well for the aAM but not the sAM; \SF is off in all cases.}
    \label{fig:chi_sp}
\end{minipage}\hfill
\begin{minipage}[t]{.482\textwidth}
\centering
\includegraphics[width=1.02\textwidth]{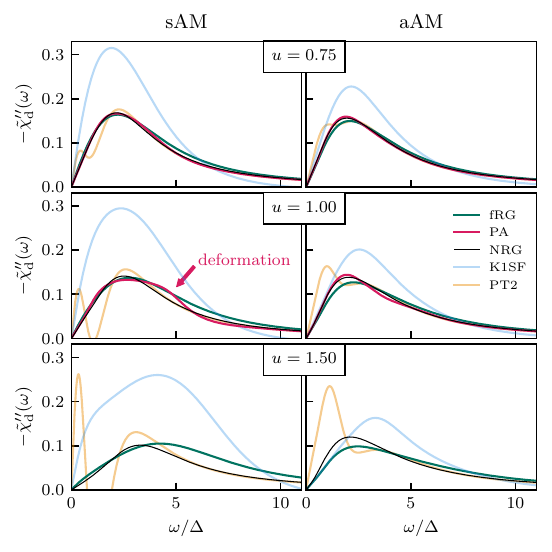}
    \caption{Imaginary part of the dynamical density susceptibility, $\tilde{\chi}_{\mathrm{d}}(\omega) \!\equiv\! \pi\Delta \chi_{\mathrm{d}}(\omega)$. Both fRG and the PA produce good results. The artifact in the PA solution at $u \!=\! 1$ in the sAM observed in Fig.~\ref{fig:self-energy} is also seen here, while it was not apparent in Fig.~\ref{fig:chi_sp}. Neither PT2 nor \SF produce sensible results for $\tilde{\chi}_{\mathrm{d}}$.
    }
    \label{fig:chi_ch}
\end{minipage}
\end{figure*}

The density susceptibility shown in Fig.~\ref{fig:chi_ch} is centered at larger frequencies and has smaller magnitude than its magnetic counterpart. Indeed, while $\tilde{\chi}_{\mathrm{m}}$ and $\tilde{\chi}_{\mathrm{d}}$ are equal at $u=0$, increasing interaction values discriminate between spin fluctuations (enhanced) and charge fluctuations (reduced).
Here, fRG and the PA both produce acceptable results. However, the PA data at intermediate $u$ and in the sAM shows a deformation around $\omega/\Delta \!\simeq\! 5$, reminiscent of the deformation in $\Sigma''$ (cf.~Fig.~\ref{fig:self-energy}). 
The K1SF curve for $\tilde{\chi}_{\mathrm{d}}$ (as for $\tilde{\chi}_{\mathrm{m}}$) is not sensible, this time lying far above (rather than below) the NRG curve. PT2 for $\chi_{\mathrm{d}}$, differently from $\chi_{\mathrm{m}}$, is unreliable, yielding a qualitatively wrong double-peak structure. 

\begin{figure}[t]
    \centering
    \includegraphics[width=0.5\textwidth]{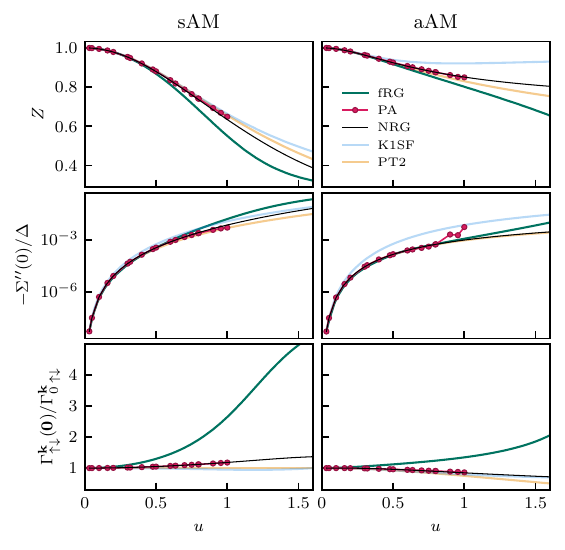}
    \caption{Static fermionic properties as a function of $u$: quasiparticle weight $Z$, scattering rate $-\Sigma''(0)$ on a logarithmic scale, and effective interaction ($\vec{k} \!=\! 12|22$) in units of the bare interaction. 
    Overall, the PA (available for $u \!\lesssim\! 1$) matches NRG best, except for $\Sigma''(0)$ at $u \!\simeq\! 1$ in the aAM. 
    All other methods agree reasonably well (except for $Z$ and $\Sigma''(0)$ in the aAM in \SF).
    Strikingly, fRG strongly overestimates the effective interaction.}
    \label{fig:QP}
\end{figure}

In summary, we find that the PA results generically reproduce the NRG benchmark best, but are available only up to intermediate $u$. Our new fRG computations with the full frequency dependence of the vertex drastically improve upon the K1SF results in almost every case, but become quantitatively off with increasing $u$.

\subsection{Static properties}\label{sec:static}
We now turn to static quantities, obtained from $\Sigma$ and $\Gamma$ by setting all frequency arguments to zero. Though these can also be obtained 
using the imaginary-frequency MF (see Ref.~\onlinecite{karrasch_SIAM_2008} for an early MF fRG treatment of the AM), they serve as important consistency checks for our Keldysh computations. The zero-frequency fermion objects can be used for an effective low-energy description, and, by rescaling, converted to quasiparticle parameters as in Hewson's renormalized perturbation theory \cite{Hewson2001}. 
For the AM in the wide-band limit at $T \!=\! 0$, the static fermionic quantities can also be deduced from the static susceptibilities.
We hence consider the static magnetic and charge susceptibilities as well, before analyzing the zero-temperature identities in the next subsection.

By virtue of the $\Delta$ flow, see Sec.~\ref{sec:AIM}, a single fRG computation suffices to obtain the \textsl{entire} dependence of, e.g., $Z(u)$ (at fixed $T/U$). By contrast, the PA requires separate computations for every value of $u$, resulting in a significantly bigger numerical effort.
The top row of Fig.~\ref{fig:QP} shows the quasiparticle weight
\begin{align}
Z
& = 
\big( 1 - \partial_\nu \Sigma' \big|_{\nu=0} \big)^{-1}, \label{eq:Z}
\end{align}
as extracted from the slope at $\nu=0$ of the real part of the retarded self-energy, $\Sigma'$.
In all cases, the PA reproduces the NRG benchmark best, but is again only available up to $u \lesssim 1$. The fRG curve follows NRG for small $u$ but starts to deviate already at intermediate $u$.
\SF performs very well in the sAM, but deviates from NRG in the aAM earlier than fRG.
Since PT2 reproduces the NRG full self-energy very well for the sAM (cf.~Fig.~\ref{fig:self-energy}), the same applies to $Z$. In the aAM, PT2 also produces reasonable results for $Z$, in contrast to $\Sigma''(\nu)$ in Fig.~\ref{fig:self-energy}.

The second row of Fig.~\ref{fig:QP} displays the scattering rate $-\Sigma''(0)$ on a logarithmic scale. In the sAM, all methods agree reasonably well up to intermediate $u$. Beyond that, fRG significantly overestimates $-\Sigma''(0)$ (cf.~Fig.~\ref{fig:self-energy}). In the aAM, the fRG results are slightly better. The PA yields the best agreement with NRG, except for $u \! \simeq \! 1$ in the aAM where numerical artifacts appear.
\SF shows large deviations early on, matching the observations in Fig.~\ref{fig:self-energy}.
PT2 reproduces NRG almost exactly, even though this is not the case for $\Sigma''(\nu)$ (Fig.~\ref{fig:self-energy}) in the aAM. 

The last row of Fig.~\ref{fig:QP} shows the effective interaction. The PA accurately reproduces the NRG results. In striking contrast,
fRG overestimates the effective interaction very strongly. (This can also be seen in Fig.~\ref{fig:fullvertex} below, third row, columns four to six, where the frequency-dependent vertex is plotted.) PT2 and \SF yield only very weak renormalizations of the bare vertex (none at all in PT2 in the sAM).

\begin{figure}
    \centering
    \includegraphics[width=0.5\textwidth]{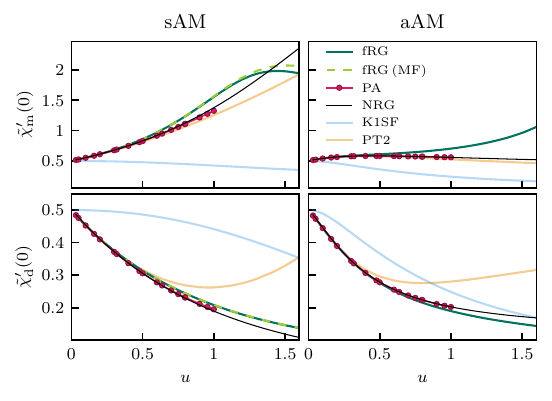}
    \caption{Static susceptibilities as a function of $u$. fRG yields sensible results until $\tilde{\chi}'_m$ has a maximum at $u \approx 1.3$. PA data are available only for $u \lesssim 1$, but show excellent agreement with NRG there. Results from \SF and PT2 (for $\tilde{\chi}_{\mathrm{d}}$) are rather bad.}
    \label{fig:stat_susc}
\end{figure}

Figure~\ref{fig:stat_susc} shows the static magnetic and density susceptibilities,
\begin{equation}
\chi_\mathrm{m} = \tfrac{1}{4} \partial_{h} \langle \tilde{n}_\uparrow - \tilde{n}_\downarrow \rangle \big|_{h=0}
, \quad
\chi_\mathrm{d} = \tfrac{1}{4} \partial_{\epsilon_d} \langle \tilde{n}_\uparrow + \tilde{n}_\downarrow \rangle
,
\end{equation}
where $\tilde{n}_\sigma  \!=\! n_\sigma \!-\! \langle n_\sigma \rangle$.
Again, the PA results, where available, reproduce the NRG benchmark best. The fRG results are reasonable up to intermediate $u$ for 
$\tilde{\chi}'_{\mathrm{m/d}}(0) \!=\! \pi\Delta \chi_{\mathrm{m/d}}$. 
A comparison with the results obtained by an independent MF computation 
(dashed lines in Figure~\ref{fig:stat_susc})
reveals that the KF data at the largest $u$ values is not fully converged in the size of the frequency grid (see App.~\ref{app:chi_m_convergence} for details).
As for the dynamical susceptibilities, \SF does not produce sensible results at all.
PT2 gives fairly good results, in particular for $\tilde{\chi}'_{\mathrm{m}}$ in the aAM (see also Fig.~\ref{fig:chi_sp}), but $\tilde{\chi}'_{\mathrm{d}}$ in the sAM quickly deviates from NRG rather strongly (as it did in Fig.~\ref{fig:chi_ch}).

In summary, for all the static properties shown in Figs.~\ref{fig:QP} and \ref{fig:stat_susc}, the PA results agree very well with NRG for all $u$ for which 
the parquet solver converged, i.e., up to $u \lesssim 1$. By contrast, fRG results begin to deviate from NRG somewhat earlier than PA, sometimes even much earlier. This difference is most striking for the effective interaction in the bottom panels of Figs.~\ref{fig:QP}, where the performance of fRG is surprisingly (even shockingly) poor.

This comparatively poor performance of fRG may be due in part to the well-known fact that one-loop fRG results depend on the choice of the fRG regulator. 
Figure~\ref{fig:Gamma000} illustrates this in the present context by comparing our KF results with independent calculations in the MF. For the latter, we used three different regulators, called $\Delta$ flow (same as for our KF computations), $U$ flow, and 
$\omega$ flow. (See Eqs.~(3) and (4) in Ref.~\onlinecite{Chalupa2022} for definitions of the $U$ and $\omega$ flow. The $\omega$ and $U$ flows require many more separate computations than the $\Delta$ flow, since the former two hold $T/\Delta$ fixed (the $\omega$ flow also $T/U$), while the latter holds $T/U$ fixed.) From Fig.~\ref{fig:Gamma000}, we note three salient points. First, the MF and KF results for the $\Delta$ flow match. This is expected for numerically converged calculations and serves as a useful consistency check. Second, the $U$ flow deviates 
from the NRG benchmark very early. 
Third,  the best MF result is obtained from the $\omega$ flow (similarly as observed in Ref.~\onlinecite{Chalupa2022}).
Regrettably, though, this advantage of the MF $\omega$ flow is not relevant for the KF: there, the $\omega$ flow would violate causality \cite{jakobs_functional_2010} and hence cannot be used. This,
and the poor performance of the $U$ flow, is the reason why we chose the $\Delta$ flow for all our KF computations.

\begin{figure}
    \centering
    \includegraphics[width=0.5\textwidth]{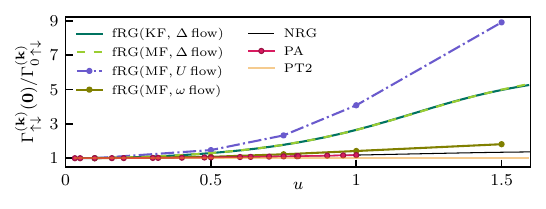}
    \caption{Effective interaction ($\vec{k} \!=\! 12|22$) of the sAM in units of the bare interaction, including fRG results in the MF obtained with three different regulators. The MF result in the $\Delta$ flow perfectly matches its KF counterpart. The $U$ flow performs considerably worse, as it quickly deviates from NRG. By far the best result is obtained using the $\omega$ flow, which can however not be used in the KF (see the main text for details).
    In the MF, we approximate vanishing frequencies by averaging over the lowest Matsubara frequencies, $\gamma_r(\vec{0})\approx \tfrac{1}{4}\sum_{\nu,\nu'=\pm\pi T} \gamma_r(0,\nu,\nu')$.
    }
    \label{fig:Gamma000}
\end{figure}

\subsection{Zero-temperature identities}\label{sec:zero-temp-identities}
As an internal consistency check for each method, we  
consider four Fermi-liquid identities. 
These hold $T \!=\! 0$ and, more generally, at $T \!\ll\! T_\mK$,
where $T_\mK$ is the Kondo temperature.
We deduce $T_\mK$ as $T_\mK \!=\! 1/[4 \chi'_\mathrm{m}(0)] |_{T=0}$ (see, e.g., Eq.~(20) in Ref.~\onlinecite{filippone_at_2018}) from zero-temperature NRG calculations. The resulting values for $u \!\in\! \{0.75, 1, 1.5\}$ are $T_\mathrm{K}/U \!\in\! \{0.31, 0.18, 0.07\}$ for the sAM and $T_\mathrm{K}/U \!\in\! \{0.58, 0.45, 0.32\}$ for the aAM. Note that the Kondo regime of the sAM corresponds to $u \!\gtrsim\! 2$ \cite{Hewson2001}.

First, for a constant hybridization function in the wide-band limit, we have the following two ``Yamada--Yosida (YY) identities'' generalized to arbitrary $\epsilon_d$
(see Eq.~(6.1) in Ref.~\onlinecite{yamada_perturbation_1975} and Eq.~(7) in Ref.~\onlinecite{yamada_perturbation_1975-1}, Eqs.~(24)--(25) in Ref.~\onlinecite{Hewson2001}, or Eqs.~(4.30)--(4.33) in Ref.~\onlinecite{kopietz_ward_2010}): 
\begin{subequations}
\begin{align}
 Z^{-1} & = [\chi_{\mathrm{m}}(0) + \chi_{\mathrm{d}}(0)]/\rho(0)
 ,
 \label{eq:YamadaYosida1}
 \\
 -\rho(0) \Gamma_{\uparrow\downarrow}(\vec{0}) & = [\chi_{\mathrm{m}}(0) - \chi_{\mathrm{d}}(0)]/\rho(0)
 .
 \label{eq:YamadaYosida2}
\end{align}
\end{subequations}
Here, $\rho(0) \!\equiv\! A(0) |_{T=0}$ is the spectral function evaluated at $\nu\!=\! 0$ and $T \!=\! 0$, 
\begin{align}
\rho(0) &= \frac{1}{\pi} \frac{\Delta}{[\epsilon_d+\Sigma'(0)]^2+\Delta^2}
\nonumber \\
&= \frac{1}{\pi\Delta} \cdot
\begin{cases}
1 & \text{for } \epsilon_d = -U/2\\
\frac{1}{1+[\Sigma'(0)/\Delta]^2} & \text{for } \epsilon_d = 0
\end{cases}
.
\end{align}
Next, $\Gamma_{\uparrow\downarrow}(\vec{0})$ is the full Matsubara 4p vertex evaluated at vanishing frequencies (in the zero-temperature limit).
The minus sign in Eq.~\eqref{eq:YamadaYosida2} stems from our convention of identifying, e.g., the bare Matsubara vertex $\Gamma_{0,\uparrow\downarrow}$ with $-U$.
The analytic continuation of $\ell$p functions between Matsubara and retarded Keldysh components involves a factor $2^{\ell/2-1}$ (see Eq.~(69) in Ref.~\onlinecite{Kugler2021}).
Hence, 
\begin{align}
\Gamma_{\uparrow\downarrow}(\vec{0})
& =
2 \Gamma^\vec{k}_{\uparrow\downarrow}(\vec{0})
,
\nonumber
\\
\vec{k} 
& \in 
\{ (12|22), (21|22), (22|12), (22|21) \} 
.
\end{align}

Another identity derived by YY (see Eqs.~(13)--(15) and (18) in Ref.~\onlinecite{yamada_perturbation_1975-1},
Eqs.~(31) and (34) in Ref.~\onlinecite{Hewson2001}, or Eq.~(4.37) in Ref.~\onlinecite{kopietz_ward_2010}) implies
\begin{align}
-\Sigma''(\nu)
& =
\tfrac{1}{2} \pi\rho(0)^3 [\Gamma_{\uparrow\downarrow}(\vec{0})]^2
(\nu^2 \!+\! \pi^2T^2)
\label{eq:Hewson}
\end{align}
for $|\nu|, T \!\ll\! T_\mK$.
We check this relation by fitting $\Sigma'' \!\propto\! (\nu^2 \!+\! \pi^2 T^2)$.
Finally, the Korringa--Shiba (KS) identity 
(see Eq.~(1.4) in Ref.~\onlinecite{Shiba1975}) reads
\begin{align}
 \lim_{\omega \rightarrow 0} \chi''_{\mathrm{m}}(\omega) / \omega 
 & =
 2\pi \left[\chi'_{\mathrm{m}}(0)\right]^2
 .
\label{eq:KorringaShiba}
\end{align}

\begin{figure}
    \centering
    \includegraphics[width=0.5\textwidth]{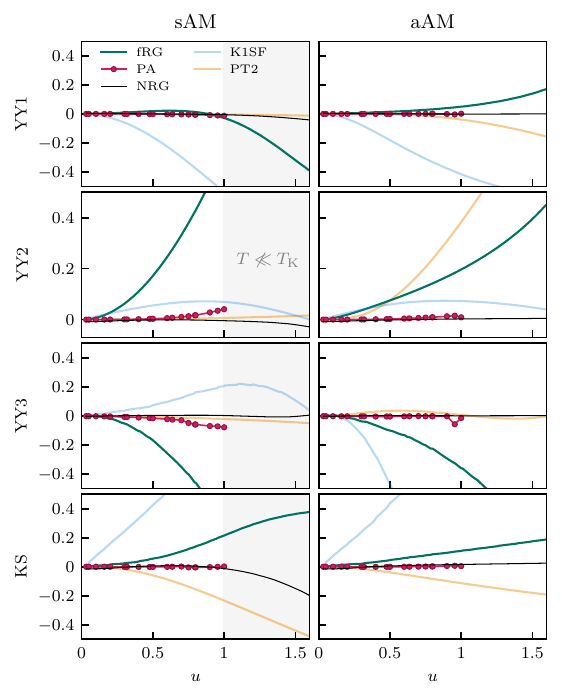}
    \caption{
    Relative difference between the left- and right-hand sides of the four zero-temperature identities as a function of $u$. All calculations have finite $T/U \!=\! 0.01$; thus, even NRG violates the identities if $T \!\ll\! T_\mathrm{K}$ is no longer fulfilled. Apart from NRG, the PA shows the smallest violations of these identities (below 8\% throughout), but is only available for $u\lesssim 1$. The fRG data fulfills YY1 relatively well, but shows clear deviations otherwise, setting in already for very small values of $u$. For YY2, e.g, the deviations become
    significant already at $u \gtrsim 0.25$.  PT2 obeys the identities (except KS) in the sAM but not the aAM.
    \SF shows major deviations throughout.}
    \label{fig:zero-T}
\end{figure}

To check the fulfillment of these identities,
we analyze the relative difference $2(\mathrm{LHS} \!-\! \mathrm{RHS})/(\mathrm{LHS} \!+\! \mathrm{RHS})$ of Eqs.~\eqref{eq:YamadaYosida1}, \eqref{eq:YamadaYosida2}, \eqref{eq:Hewson} \eqref{eq:KorringaShiba}, referred to as YY1, YY2, YY3, KS, respectively.
These zero-temperature identities of the AM only hold if $T \ll T_\mathrm{K}$. As we keep $T/U \!=\! 0.01$ constant, the temperatures increase with $u$, and $T \!\ll\! T_\mathrm{K}$ is no longer fulfilled for $u \!\gtrsim\! 1$ in the sAM. Accordingly, there, the identities are violated even in NRG. 

As can be seen in Fig.~\ref{fig:zero-T}, the PA fulfills most identities very well (below 8\% throughout), but is again available only up to $u \!\simeq\! 1$. 
The fRG results obey YY1 up to $u \!\lesssim\! 1$, but show clear deviations in all other identities, setting in already for for very small values of $u$.
Except for the KS relation in the fourth row, PT2 mostly fulfills the identities for the sAM but less so for the aAM, while \SF shows major deviations, even for small $u$. 

\subsection{Frequency dependence of the 4p vertex}\label{sec:vertex}
\begin{figure*}
    \centering
    \includegraphics[width=\textwidth]{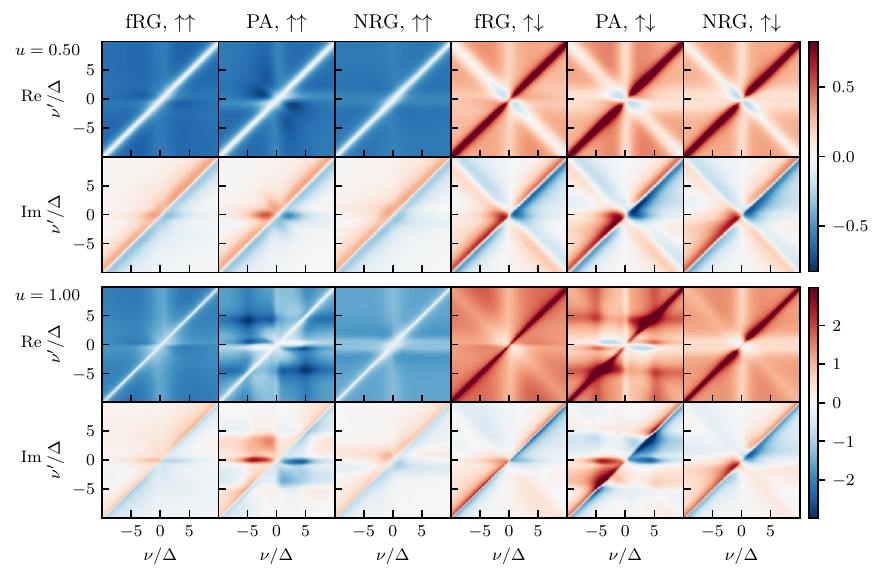}
    \caption{Fully retarded ($\vec{k}\!=\!12|22$) Keldysh component of the full vertex, $(\Gamma^{\vec{k}}_{\sigma\sigma'}(\omega_t \!=\! 0, \nu_t \!=\! \nu, \nu_t' \!=\! \nu') - \Gamma^{\vec{k}}_{0,\sigma\sigma'})/\Gamma^{\vec{k}}_{0\, \uparrow\downarrow}$, for $u \!=\! 0.5$ (top panel) and $u \!=\! 1$ (bottom), computed using fRG, PA and NRG (following Refs.~\cite{Kugler2021,Lee2021}). 
    We observe very good agreement for $u \!=\! 0.5$, which, qualitatively, mostly persists for higher interaction. However, $\mathrm{Re}\,\Gamma_{\uparrow\downarrow}$ at $u \!=\! 1$ and low frequencies differs significantly between the methods: it is strictly positive in fRG, slightly negative in NRG, but much more strongly negative up to fairly large values of $\nu$ in the PA.
    Generally, the PA shows more complicated features than NRG for larger $u$, despite being numerically converged, indicating the breakdown of the PA.}
    \label{fig:fullvertex}
\end{figure*}
Finally, we show fRG and PA results for the frequency dependence of the 4p vertex in the sAM and compare them to corresponding results from NRG. We restrict ourselves to a fully retarded Keldysh component \cite{Kugler2021} and show both the same-spin ($\uparrow\uparrow$) and the opposite-spin ($\uparrow\downarrow$) components. We plot the vertex in the two-dimensional frequency plane $(\omega_t \!=\! 0, \nu_t \!=\! \nu, \nu_t' \!=\! \nu')$ in the natural parametrization of the $t$ channel for zero transfer frequency. Physically, this corresponds to the effective interaction of two electrons on the impurity with equal or opposite spins, respectively, and energies $\nu, \nu'$ without energy transfer \cite{kopietz_introduction_2010}. The NRG 4p results are computed with the scheme introduced in Refs.~\onlinecite{Kugler2021, Lee2021},
utilizing the symmetric improved estimator of Ref.~\onlinecite{Lihm2023}.

In Fig.~\ref{fig:fullvertex}, we compare results from fRG, the PA, and NRG for two values of the interaction $u \in \{0.5, 1\}$. We observe good qualitative agreement throughout, as all methods capture all nontrivial features. 
At $u \!=\! 1$, however, we observe a qualitative discrepancy in the data:
$\mathrm{Re}\,\Gamma_{\uparrow\downarrow}$ is strictly positive in fRG and slightly negative in NRG 
(bottom part, top row, first panel from the right in Fig.~\ref{fig:fullvertex}). 
The PA result reaches even larger negative values and retains them for a large range of $\nu$ values.
This strong negative signal appears to be an artifact of the PA; it
would likely be canceled by additional contributions missed in the PA.

\section{Conclusions and outlook}\label{sec:Conclusion}
In this work, we have shown that real-frequency QFT calculations with full frequency resolution of the 4p vertex \textsl{are} feasible. We chose the AM for a proof-of-principle study and employed one-loop 
fRG flows and solutions of the parquet equations in the PA, benchmarked against NRG. We compared dynamical correlation functions as well as characteristic static quantities and performed a detailed numerical check of zero-temperature identities. We found that keeping the full frequency dependence of the 4p vertex in fRG strongly improves the accuracy compared to previous implementations using functions with at most one-dimensional frequency dependencies. Note that the present study is performed at finite temperature, $T/U \!=\! 0.01$, in contrast to previous work on spectral functions at $T \!=\! 0$ \cite{jakobs_nonequilibrium_2010}.

The numerical challenges imposed by the fully parametrized real-frequency 4p vertex were overcome via a suitably adapted frequency grid, vectorization over the Keldysh matrix structure, and a parallelized evaluation of the fRG or parquet equations (see App.~\ref{app:implementation}). We employed frequency grids with up to $125^3$ data points, and our most expensive calculation consumed about 25000 CPU hours for a single data point in the PA.

The PA results could be converged only for
$u \!=\! U/(\pi\Delta)$ in the range $u \lesssim 1$, but there gave the best agreement with NRG (except at the boundary of the accessible $u$ range). The PA also gave very good results for the effective interaction. However, by looking at $\Gamma^\vec{k}_{\uparrow\downarrow}$ in a frequency range around the origin, it appears that the mechanism by which the PA achieves low values of $|\Gamma^\vec{k}_{\uparrow\downarrow}(\vec{0})|$ (compared to, say, fRG) is different from that of NRG, as the PA data has 
a spuriously large regime of strongly negative values in $\mathrm{Re}\,\Gamma^\vec{k}_{\uparrow\downarrow}$.

The fRG calculations in the present context were comparatively economical, since a single run with the ``$\Delta$ flow'' yields an entire parameter sweep in $\Delta$. The flow could be followed to large values of $u$, well beyond $1$, i.e., far beyond the regime where we could converge the PA. However, for $u \gtrsim 0.5$ these one-loop fRG results are significantly less accurate than the PA (as compared to NRG). Strikingly, fRG strongly overestimates the effective interaction $\Gamma^\vec{k}_{\uparrow\downarrow}(\vec{0})$ by factors of $3$ to $4$ for $u$ in the range $1$ to $1.5$. We compared the Keldysh to Matsubara fRG data obtained using three different regulators, and we found that, for $u \!>\! 0.5$, the latter strongly depend on the choice of regulator: For the $\Delta$ flow, the Matsubara results agree with the Keldysh results, while performing better than the $U$ flow but worse than the $\omega$ flow.  
Regrettably, the $\omega$ flow is not available in the KF, where it violates causality. It would hence be worthwhile to find Keldysh fRG regulators akin to the $\omega$ flow but compatible with the KF requirements regarding causality and FDTs \cite{jakobs_functional_2010}.

The regulator dependence in fRG can be eliminated in the multiloop fRG framework, yielding results equivalent to the PA upon convergence in the number of loops \cite{kugler_multiloop_2018,kugler_multiloop_2018-1,kugler_derivation_2018}. 
This has been demonstrated numerically in imaginary frequencies for the AM \cite{Chalupa2022} 
(and in Refs.~\onlinecite{tagliavini_multiloop_2019,hille_quantitative_frg_2020} for the Hubbard model).
Yet, using a multiloop extension of our Keldysh fRG code, we found the computation of multiloop contributions considerably harder for Keldysh vertices than for Matsubara vertices.
The reason seems to be that, for real-frequency Keldysh vertices, the higher-loop contributions for increasing $u$ show a considerably more complicated frequency structure than the original fRG vertex itself (similarly to how the PA vertex has more structure than its fRG counterpart in the bottom panel of Fig.~\ref{fig:fullvertex}).
A more detailed analysis along these lines is however left for future work.

Our work paves the way for many follow-up studies.
For instance, one can exploit the power of the KF to study non-equilibrium phenomena, and the AM with a finite bias voltage is tractable with only minor increase in the numerical costs \cite{Fujii2003,jakobs_nonequilibrium_2010}. 
Further, we here considered moderate interaction strengths $u \!\lesssim\! 1.5$
as it is known that fRG and the PA are unable to access the non-perturbative regime of the AM \cite{Chalupa2021,Chalupa2022} or, e.g., the Hubbard model \cite{hille_quantitative_frg_2020,Eckhardt2020}.
An important future direction is, therefore, to use these methods in a more indirect manner, as real-frequency diagrammatic extensions \cite{Rohringer2018} of dynamical mean-field theory \cite{Georges1996}. The first, established building block for this is the non-perturbative input, namely 2p and 4p vertices, from NRG \cite{Kugler2021,Lee2021}. The present work presents another building block: real-frequency QFT with full frequency resolution of the 4p vertex. 
An important next step will be to use our diagrammatic framework to study the consistency of the NRG results for the 2p and 4p vertices,  e.g., by checking whether they fulfill the SDE. The final building block will then be to include momentum degrees of freedom in real-frequency QFT approaches built on top of NRG. 

Keeping track of the momentum dependence will lead to a major increase in numerical complexity. This can be addressed using economical implementations and compression algorithms such as truncated-unity approaches \cite{husemann_efficient_2009, wang_functional_2012, lichtenstein_high-performance_2017, eckhardt_truncated-unity_2018} or the new quantics tensor cross interpolation scheme \cite{nunez_fernandez_learning_2022, Shinaoka2023, ritter_quantics_2023}. 
The latter can be used to obtain highly compressed tensor network representations of multi-dimensional functions, 
potentially leading to exponential reductions in computational costs.
First investigations have shown that the objects encountered in diagrammatic many-body approaches may indeed have strongly compressible quantics representations \cite{Shinaoka2023}.

\section*{Data and Code Availability}
All raw data required to reproduce the plots as well as the full data analysis and the plotting scripts are available under \url{https://opendata.physik.lmu.de/ar879YbJwUpAM2S}. A separate publication of the fully documented source code used to generate the raw data is in preparation.

\acknowledgements
We thank Jeongmin Shim and Seung-Sup B.~Lee for providing data of the NRG 4p vertex as well as Severin Jakobs for a careful reading of the manuscript.
The numerical simulations were performed on the Linux clusters and the SuperMUC cluster (project 23769) at the Leibniz Supercomputing Center in Munich.
NR acknowledges funding from a graduate scholarship from the German Academic Scholarship Foundation (Studienstiftung des deutschen Volkes) and additional support from the ``Marianne-Plehn-Programm'' of the state of Bavaria. AG and JvD were supported by the Deutsche Forschungsgemeinschaft under Germany’s Excellence Strategy EXC-2111 (Project No.~390814868), and the Munich Quantum Valley, supported by the Bavarian state government with funds from the Hightech Agenda Bayern Plus. FBK acknowledges support by the Alexander von Humboldt Foundation through the Feodor Lynen Fellowship. The Flatiron Institute is a division of the Simons Foundation.  

\appendix

\section{The two-particle vertex}
\label{app:vertex_parametrization}

\begin{figure*}[t]
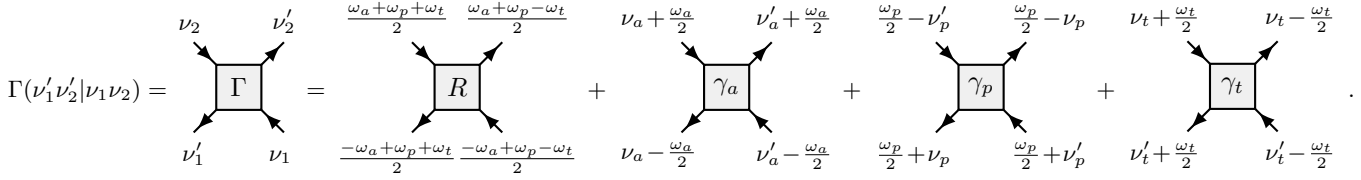

    \centering
    \begin{alignat}{5}
	\Gamma(\nu_1'\nu_2'|\nu_1\nu_2)
	\nonumber
	&= \!\!
	\tikzinclude{Keldysh_vertex_parametrization-channeldec-PE}
	\!\! &&= \!\!
	\tikzinclude{Keldysh_vertex_parametrization-channeldec-PE_R_bosonic}
	\! &&+ \!
	\tikzinclude{Keldysh_vertex_parametrization-channeldec-PE_a}
	\! &&+ \!
	\tikzinclude{Keldysh_vertex_parametrization-channeldec-PE_p}
	\! &&+ \!
	\tikzinclude{Keldysh_vertex_parametrization-channeldec-PE_t}
	\nonumber 
	\,.
\end{alignat}
    \caption{We show the frequency conventions for the two-particle reducible vertices $\gamma_r$ with $r=a,p,t$. Symmetric parametrizations with $\pm \tfrac{\omega}{2}$ ensure that vertex structures are centered around the frequency axis. The irreducible vertex $R$ is shown in bosonic frequencies for completeness.}
    \label{fig:parametrization:frequency_convention}
\end{figure*}

In compact notation, we denote the vertex by $\Gamma_{1'2'|12}$ where each leg carries a multi-index $i= (k_i, \sigma_i, \nu_i )$ with Keldysh index $k_i$, spin $\sigma_i$, and fermionic frequency $\nu_i$.
Generic symmetries of the full Keldysh vertex are derived in Ref.~\onlinecite{jakobs_properties_2010} and other symmetries (such as spin or particle-hole symmetry) are given in Ref.~\onlinecite{rohringer_local_2012}.
In the following, we recap these symmetries and detail the parametrization in our implementation.
First, we work with Keldysh indices rather than contour indices. 
In this basis, the $11\ndots 1$ ($22\ndots 2$) component of a multipoint correlator (vertex) vanishes \cite{jakobs_properties_2010}.
This simplifies, e.g., the Dyson equation, $G^R = [(G_0^R)^{-1} - \Sigma^R]^{-1}$ and implies $\Gamma^{22|22}=0$.
Furthermore, crossing symmetry gives 
\begin{align}
    \Gamma_{1'2'|12} = -\Gamma_{2'1'|12} = -\Gamma_{1'2'|21} = \Gamma_{2'1'|21},
    \label{eq:vertex_symmetries_and_parametrization:crossing_symmetry}
\end{align}
and complex conjugation
\begin{align}
    \Gamma_{1'2'|12} = (-1)^{1+\sum_i k_i}\Gamma_{12|1'2'}^*.
    \label{eq:vertex_symmetries_and_parametrization:complex_conjugation}
\end{align}
Thermal equilibrium entails (generalized) fluctuation-dissipation relations between different Keldysh components. However, we choose to vectorize the code over Keldysh components and thus do not use these relations (see App.~\ref{app:implementation} for details on the vectorization).
For a comprehensive list of multi-point fluctuation-dissipation relations, we refer to Refs.~\onlinecite{wang_generalized_2002,jakobs_properties_2010,halbinger_2023}.
They are very well fulfilled (percent level) by our numerical results.

In the absence of a magnetic field, spin conservation and the invariance under a global spin flip reduce the number of independent spin components.
The remaining components are related by the SU(2) relation \cite{rohringer_local_2012} 
\begin{align}
\label{eq:su2spin_relation}
\Gamma_{\sigma\sigma|\sigma\sigma} = \Gamma_{\sigma\bar\sigma|\sigma\bar\sigma} + \Gamma_{\sigma\bar\sigma|\bar\sigma\sigma}    
\end{align}
where $\bar{\uparrow} = \downarrow$ and vice versa.
Hence, the spin dependence of the vertex can be parametrized by
\begin{align}
    \label{eq:vertex_spin_parametrization}
    \Gamma_{\sigma_1'\sigma_2'|\sigma_1\sigma_2} &= 
    \Gamma_\updown    \delta_{\sigma_1',\sigma_1}\delta_{\sigma_2',\sigma_2}
+   \Gamma_\updownbar \delta_{\sigma_1',\sigma_2}\delta_{\sigma_2',\sigma_1}.
\end{align}
The components on the right-hand side are related by crossing symmetry. It thus suffices to compute a single one of them.
At particle-hole symmetry, we further have 
\begin{align}
\nonumber
    &\Gamma_{1'2'|12}(\nu_{1}',\nu_{2}'|\nu_{1},\nu_{2}) = \Gamma_{12|1'2'}(-\nu_{1},-\nu_{2}|-\nu_{1}',-\nu_{2}') 
    \\
    &
    \overset{\eqref{eq:vertex_symmetries_and_parametrization:complex_conjugation}}= 
    (-1)^{1+\sum_i k_i}\Gamma_{1'2'|12}(-\nu_{1}',-\nu_{2}'|-\nu_{1},-\nu_{2})^*
        \label{eq:vertex_symmetries_and_parametrization:particle_hole_symmetry}
\end{align}
with the multi-indices $i=(k_i,\sigma_i)$, reducing the number of independent frequency components even more.

By frequency conservation, $\nu_{1}' \!+\! \nu_{2}' \!=\! \nu_{1} \!+\! \nu_{2}$, the vertex depends on only three independent frequencies. 
These are chosen differently for each two-particle reducible vertex $\gamma_r$ (see Fig.~\ref{fig:parametrization:frequency_convention}), with the bosonic transfer frequency $\omega_r$ and the fermionic frequencies $\nu_r$ and $\nu_r'$. 
The vertices $\gamma_r$ have non-trivial asymptotics in the limits $\nu_r^{(\prime)} \!\to\! \infty$.
One can decompose the reducible vertex $\gamma_r$ in asymptotic classes, see Eq.~\eqref{eq:asymptotic_decomposition} \cite{wentzell_high-frequency_2020}.
Since the bare interaction is frequency independent, the asymptotic classes $\Ktot{ir}$ can be identified with certain diagrams that are reducible in channel $r$ \cite{wentzell_high-frequency_2020,thoenniss_multiloop_2020}. Connecting two external legs to the same bare interaction vertex reduces the dependence by one external frequency argument.
$\Ktot{1r}(\omega_r)$ consists of all diagrams where the two external legs carrying frequency $\nu_r$ connect to the same bare vertex and the external legs carrying $\nu'_r$ connect to another one. Hence, $\Ktot{1r}$ only depends on $\omega_r$.
$\Ktot{2r}(\omega_r,\nu_r)$ consists of all diagrams where the $\nu_r'$ legs connect to the same bare vertex while each of the other two legs connect to different bare vertices.
$\Ktot{2'r}(\omega_r,\nu'_r)$ is analogous to $\Ktot{2r}$ with the roles of $\nu_r$ and $\nu_r'$ interchanged.
For $\Ktot{3r}(\omega_r,\nu_r,\nu_r')$ all external legs connect to different bare vertices.
 
The bare vertices simplify not only the dependence of $\Ktot{1}$, $\Ktot{2}$, and $\Ktot{2'}$ on frequencie but also on Keldysh indices.
If a bare vertex connects to two external legs, flipping their Keldysh indices, $\bar{1} \!=\! 2$ ($\bar{2} \!=\! 1$), leaves the function invariant, see Eq.~\eqref{eq:Keldysh_formalism:Field_theory:2P:Bare_vertex_Keldysh_indices}.
This gives, e.g.,
\begin{subequations}
\begin{align}
\nonumber
    K_{1p}^{k_{1'}k_{2'}|k_{1}k_{2}} 
    &=
    K_{1p}^{\bar{k}_{1'}\bar{k}_{2'}|k_{1}k_{2}} 
    =    
    K_{1p}^{k_{1'}k_{2'}|\bar{k}_{1}\bar{k}_{2}} 
    \\&=
     K_{1p}^{\bar{k}_{1'}\bar{k}_{2'}|\bar{k}_{1}\bar{k}_{2}}
    ,
    \\
    \Ktot{2p, \sigma_{1'}\sigma_{2'}|\sigma_{1}\sigma_{2}}^{k_{1'}k_{2'}|k_{1}k_{2}}
    &= 
    \Ktot{2p, \sigma_{1'}\sigma_{2'}|\sigma_{1}\sigma_{2}}^{k_{1'}k_{2'}|\bar{k}_{1}\bar{k}_{2}} 
    .
\end{align}
\end{subequations}
Note that the diagrammatic channels $a$ and $t$ flip under crossing symmetry, \ie $\gamma_{a, 1'2'|12} = -\gamma_{t, 1'2'|21}$, while channel $p$ is crossing symmetric itself.
The symmetry relations in Eqs.~\eqref{eq:vertex_symmetries_and_parametrization:crossing_symmetry}--\eqref{eq:vertex_symmetries_and_parametrization:particle_hole_symmetry} are formulated for full vertices. They can be adapted to the asymptotic classes $\Ktot{ir}$ by inserting the decomposition on both sides of each relation and taking the appropriate limits $\nu_r^{(\prime)}\!\to\!\infty$.
For instance, $\Ktot{\updown,2'p}$ is related to $\Ktot{\updown,2p}$ by 
\begin{align}
    \Ktot{\updown,2'p}^{k_{1'}k_{2'}|k_{1}k_{2}}(\omega_p,\nu_p') & \overset{\eqref{eq:vertex_symmetries_and_parametrization:complex_conjugation}}{=}
     (-1)^{1+\sum_i{k_i}} \Ktot{\updown,2p}^{k_{1}k_{2}|k_{1'}k_{2'}}(\omega_p,\nu_p')
     \,.
\end{align}
For a formulation of the parquet and fRG equations in terms of asymptotic classes, we refer to Ref.~\onlinecite{wentzell_high-frequency_2020} and to Eqs.~(75) in Ref.~\onlinecite{gievers_multiloop_2022}.

As we vectorize over Keldysh indices, we explicitly keep track of all Keldysh components. 
The symmetry relations are then used to reduce the spin and frequency components
(Eqs.~\eqref{eq:vertex_symmetries_and_parametrization:crossing_symmetry}, \eqref{eq:vertex_symmetries_and_parametrization:complex_conjugation}, and \eqref{eq:vertex_symmetries_and_parametrization:particle_hole_symmetry}
for $\Gamma_{\updown}$).
To implement these symmetries for the $K_{3r}$ class, it is convenient to express the relations in terms of the three bosonic frequencies \cite{karrasch_SIAM_2008}, giving
\begin{align}
    \Gamma^{k_{1'}k_{2'}|k_{1}k_{2}}_{\updown; \omega_a,\omega_p,\omega_t}
    \nonumber
    \overset{\eqref{eq:vertex_symmetries_and_parametrization:complex_conjugation}}{=}&
    \big[
    \Gamma^{k_{1}k_{2}|k_{1'}k_{2'}}_{\updown;\, \omega_a,\omega_p,-\omega_t}
    \big]^* (-1)^{1+\sum_i k_i}
    \\
    \overset{\eqref{eq:vertex_symmetries_and_parametrization:crossing_symmetry}}{=}
    \Gamma^{k_{2'}k_{1'}|k_{2}k_{1}}_{\updown;\, -\omega_a,\omega_p,-\omega_t}
    \overset{\eqref{eq:vertex_symmetries_and_parametrization:particle_hole_symmetry}}{=}&
    \big[
    \Gamma^{k_{1'}k_{2'}|k_{1}k_{2}}_{\updown;\, -\omega_a,-\omega_p,-\omega_t}
    \big]^* (-1)^{1+\sum_i k_i}
    ,
\end{align}
such that the sign of the bosonic frequencies define sectors that are related by symmetry.

\section{Frequency dependence of vertex components}\label{app:vertex-components}
\begin{figure*}[ht]
    \centering
    \parbox{0.49\textwidth}{
        \includegraphics[width=0.49\textwidth]{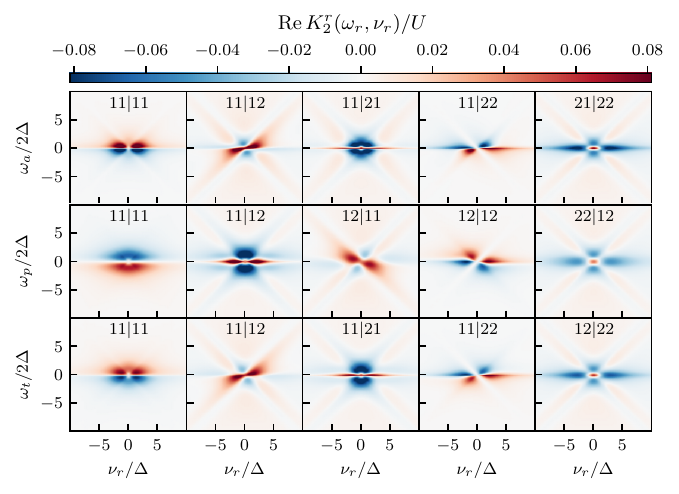}
        \vspace{-5 ex}
        \caption*{\hspace{55 ex} $\boxed{u=0.5}$}
    }
    \parbox{0.49\textwidth}{
        \includegraphics[width=0.49\textwidth]{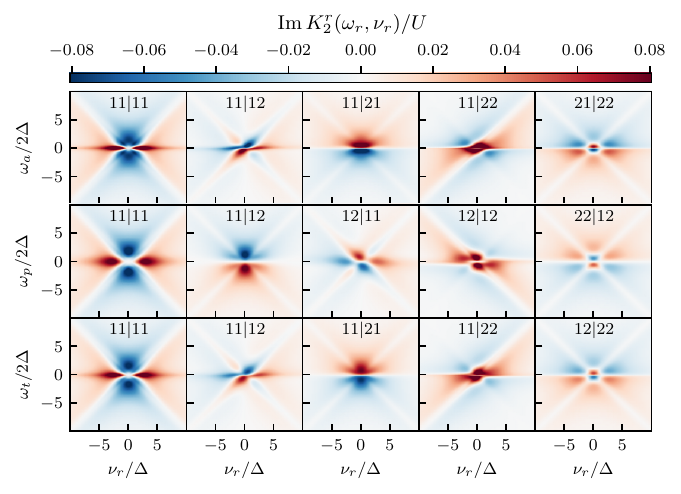}
        \vspace{-5 ex}
        \caption*{\phantom{$\boxed{u=0.5}$}}
    }
    \parbox{0.49\textwidth}{
        \vspace{-1ex}
        \includegraphics[width=0.49\textwidth]{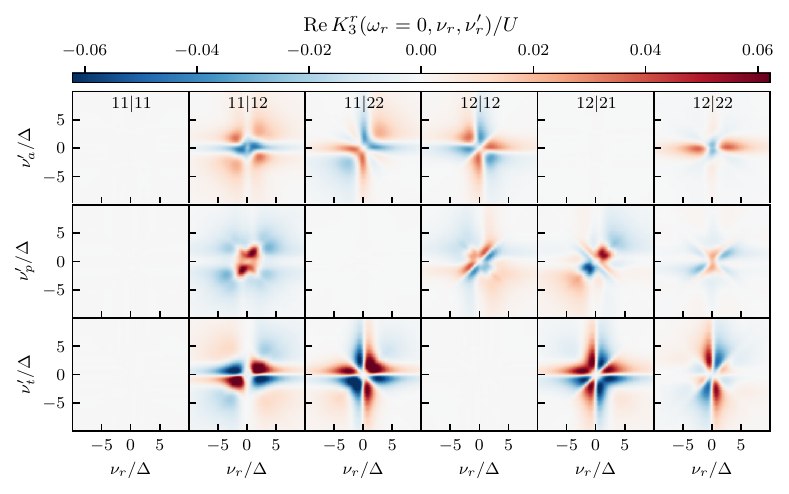}
    }
    \parbox{0.49\textwidth}{
        \vspace{-1ex}
        \includegraphics[width=0.49\textwidth]{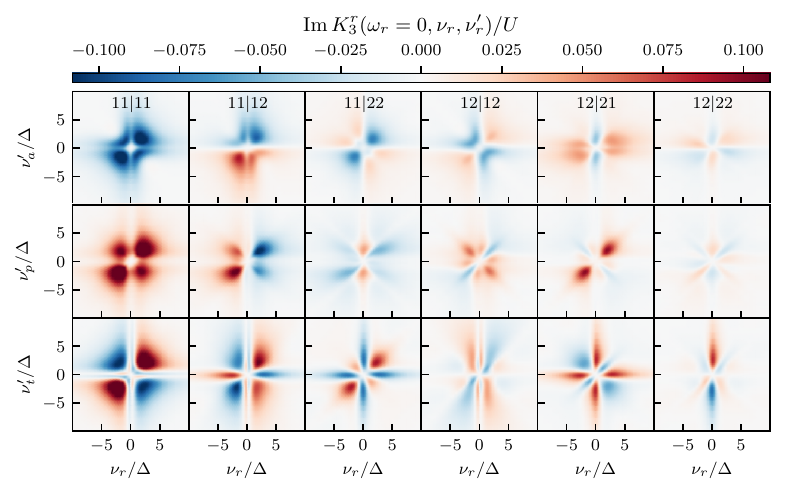}
    }
    \caption{Real (left) and imaginary (right) parts of $K_2$ (top) and $K_3$ (bottom) vertex components in the PA for $u = 0.5$. The three rows of each subfigure show results for the three two-particle channels $r \in \{a,p,t\}$. The columns show all independent Keldysh components. Natural frequency parametrizations were used and for $K_3$ the bosonic transfer frequency $\omega_r$ was set to zero. Consequently, some components of $\mathrm{Re}\, K_3$ vanish.
    }
    \label{fig:vertex_0.5}
\end{figure*}
\begin{figure*}[ht]
    \centering
    \parbox{0.49\textwidth}{
        \includegraphics[width=0.49\textwidth]{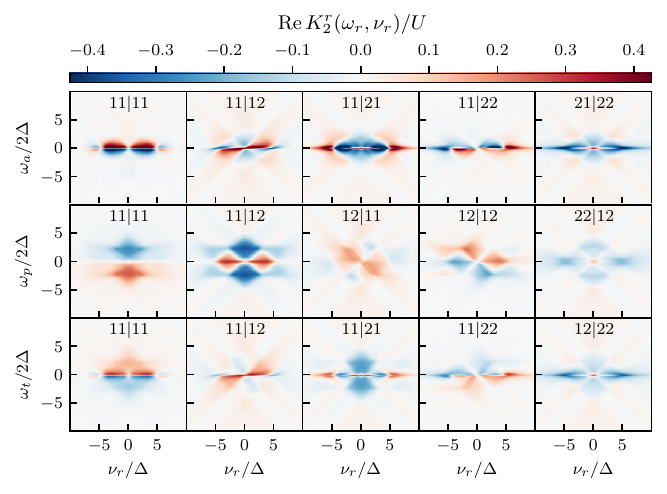}
        \vspace{-5 ex}
        \caption*{ \hspace{55 ex} $\boxed{u=1.0}$}
    }
    \parbox{0.49\textwidth}{
        \includegraphics[width=0.49\textwidth]{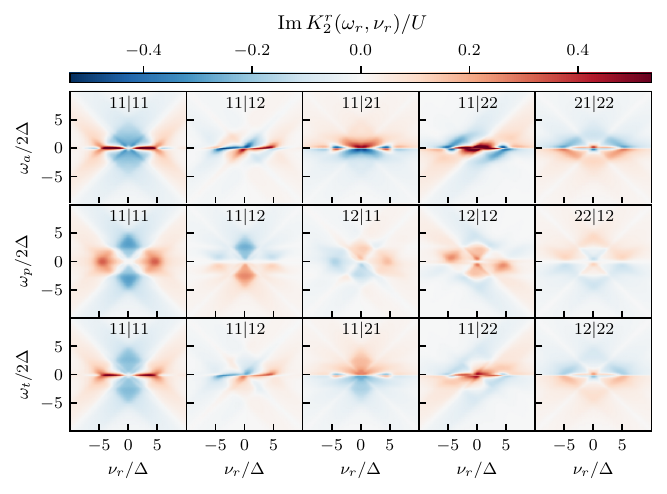}
        \vspace{-5 ex}
        \caption*{\phantom{$\boxed{u=1.0}$}}
    }
    \parbox{0.49\textwidth}{
        \vspace{-1 ex}
        \includegraphics[width=0.49\textwidth]{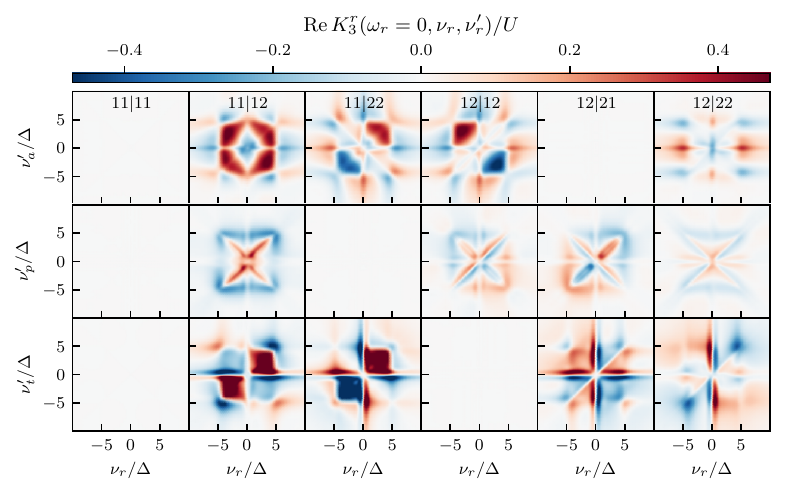}
    }
    \parbox{0.49\textwidth}{
        \vspace{-1 ex}
        \includegraphics[width=0.49\textwidth]{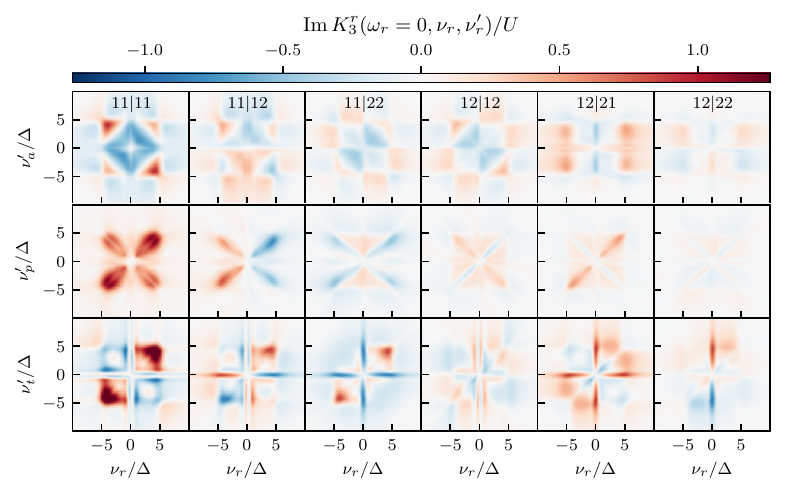}
    }
    \caption{Same vertex components as in Fig.~\ref{fig:vertex_0.5}, computed in the PA for $u = 1$.}
    \label{fig:vertex_2.5}
\end{figure*}
Figures~\ref{fig:vertex_0.5} and \ref{fig:vertex_2.5} show plots for the frequency dependence of the asymptotic classes $K_2$ and $K_3$ for each of the three two-particle channels $r \!\in\! \{a,p,t\}$, computed in the PA for $u \!=\! 0.5$ and $u \!=\! 1$, respectively. 
We use the natural frequency parametrization for each channel $r$ and set the bosonic transfer frequency $\omega_r \!=\! 0$ in the plots for $K_3$. The figures instructively show 
what types of non-trivial structures emerge during such calculations. In particular, one can clearly see that the frequency resolution needs to be very 
high throughout to resolve all sharp features (many occurring on scales 
much smaller than $\Delta$). Moreover, the weak-coupling results may serve as benchmarks for future computations of Keldysh vertices using other methods, such as NRG or QMC. 

\section{Fully parametrized equations}
\label{app:Fully_Parametrized_Functions}

We can write the parquet equations~\eqref{eq:all_parquet_equations} and one-loop fRG flow equations~\eqref{eq:mfRG:fRG:Flow_equations} entirely in terms of two functions, bubbles and loops.
A bubble $B_r$ in channel ${r=a,p,t}$ combines two vertices via a propagator pair
\begin{subequations}
\label{eq:propagator_pair}
\begin{align}
    \Pi_{a, \omega_a\nu_a}^{34|3'4'} &= G^{3|3'}_{\nu_a - \omega_a/2} G^{4|4'}_{\nu_a + \omega_a/2}
    ,
    \\
    \Pi_{p, \omega_p\nu_p}^{34|3'4'} &= G^{3|3'}_{\omega_p/2 + \nu_p} G^{4|4'}_{\omega_p/2 - \nu_p}
    ,
    \\
    \Pi_{t, \omega_t\nu_t}^{43|3'4'} &= G^{3|3'}_{\nu_t - \omega_t/2} G^{4|4'}_{\nu_t + \omega_t/2}
    ,
\end{align}
\end{subequations}
where we use the natural frequency parametrization for each channel (see Fig.~\ref{fig:parametrization:fully_parametrized_bubbles}) and superscripts indicate Keldysh indices $(34|3'4') = (k_{3}k_{4}|k_{3'}k_{4'})$.
In the following, we give explicit formulas for the $\updown$-spin component of bubble diagrams that combine vertices $V$ and $W$:
\begin{subequations}
\label{eq:parametrization:bubbles}
\begin{align}
&B_a[V,W]_{\updown,\omega_a\nu_a\nu'_a}^{1'2'|12}                
    \!= 
   \!\!\int_{\tilde \nu}\!\!
	V_{\updown,\omega_a \nu_a \tilde{\nu}}^{1'4'|32}
 \Pi^{34|3'4'}_{a,\omega_a \tilde{\nu}}
   W^{3'2'|14}_{\updown,\omega_a \tilde{\nu} \nu_a'}
 ,
 \\
&B_p[V,W]_{\updown,\omega_p\nu_p\nu_p'}^{1'2'|12} \!
 =  \!\!\int_{\tilde \nu}\!\!
	 V_{\updown,\omega_p\nu_p\tilde{\nu}}^{1'2'|34}
  \Pi^{34|3'4'}_{p,\omega_p \tilde{\nu}} 
   W^{3'4'|12}_{\updown,\omega_p \tilde{\nu} \nu_p'} 
 ,
\intertext{with $\int_{\tilde{\nu}} = \int_{-\infty}^\infty\tfrac{\mathrm{d}\tilde \nu}{2\pi\mi}$ 
(the internal spin sum and crossing symmetry in $B_{p}$ cancel the prefactor of $1/2$), and}
&B_t[V,W]_{\updown,\omega_t \nu_t \nu_t'}^{1'2'|12}
    =
    -\!\!\int_{\tilde \nu}\!\!
    \Pi^{43|3'4'}_{t, \omega_t \tilde{\nu}}
\\&
    \phantom{B[V,W]}
    \times  \big[
	V_{\updown,\omega_t,\nu_t,\tilde{\nu}}^{4'2'|32}   
   W^{1'3'|14}_{\upup,\omega_t \tilde{\nu} \nu_t'}          \nonumber
 +
	V_{\upup,\omega_t\nu_t\tilde{\nu}}^{4'2'|32}
   W^{1'3'|14}_{\updown, \omega_t \tilde{\nu} \nu_t'}        \nonumber
 \big]
 ,
\end{align}
\end{subequations}
where the $\upup$-spin component is obtained via Eq.~\eqref{eq:vertex_spin_parametrization}.

For the loop, we parametrize the vertex in the $t$-channel convention with $\omega_t=0$ and write
\begin{align}
\label{eq:parametrization:loop}
    L[\Gamma,G]^{1'|1}_{\nu} =& - \!\!\int_{\tilde\nu}\!\!
     G^{2|2'}_{\nu_t}
     \left[
     \Gamma_{\updown}
     +
     \Gamma_{\upup}
     \right]^{1'2'|12}_{0 \nu_t \nu}
     .
\end{align}

Using the loop $L$ and bubbles $B_r$, the parquet equations~\eqref{eq:all_parquet_equations} read
\begin{subequations}
\begin{align}
    \gamma_r &= B_r[I_r,\Gamma]
    , \\
    \Sigma &=
    L[\Gamma_0, G] + \tfrac{1}{2} L[B_a[\Gamma_0,\Gamma],G]
    . 
    \label{eq:background:Schwinger-Dyson-equation}
\end{align}
\end{subequations}
In the SDE, the internal spin sum can be performed,
canceling the factor of $1/2$ in Eq.~\eqref{eq:background:Schwinger-Dyson-equation} by crossing symmetry, to give
\begin{subequations}
\label{eq:all_parquet_equations_as_formulas}
\begin{align}
    \Sigma^{1'|1}_{\text{SDE} \nu} =&
    -\!\!\int_{\nu_t}\!\!
     G^{2|2'}_{\nu_t}
     \left[
    \Gamma_{0, \updown}
    +
     B_a[\Gamma_0,\Gamma]^{1'2'|12}_{\updown, 0 \nu_t \nu}
     \right]
     .
\end{align}
\end{subequations}

The one-loop fRG flow equations [cf.\ Eq.~\eqref{eq:mfRG:fRG:Flow_equations}] 
are
\begin{align}
\label{eq:mfRG:fRG:Flow_equations_as_formulas}
\dot \Sigma &= L(\Gamma, S)
, \qquad
\dot{\gamma}_r = \dot{B}_r(\Gamma,\Gamma) 
,
\end{align}
where the dot on $\dot{B}_r$ denotes a differentiated propagator pair, $\partial_\Lambda {\Pi}_r = \dot G G + G \dot G$, including the Katanin substitution $S \to \dot G = S + G \dot \Sigma G$ \cite{katanin_fulfillment_2004}.

\begin{figure}
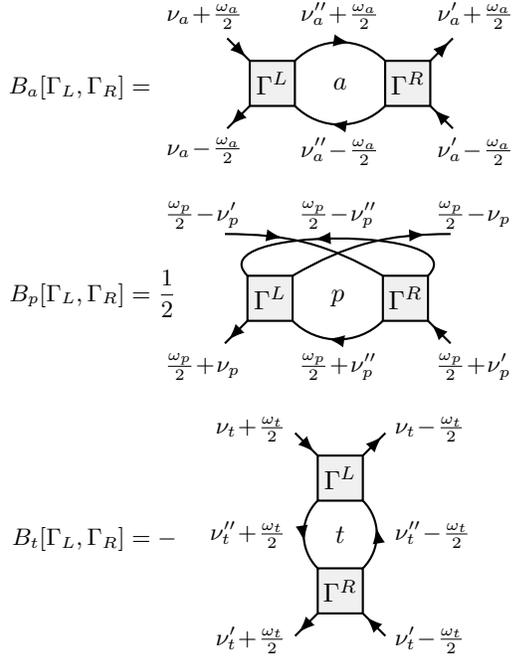

    \centering
    \begin{align*}
    B_a[\Gamma_L,\Gamma_R]
    &=
	\tikzinclude{Keldysh_vertex_parametrization-channeldec-Ba}
	\!\!\!
    \\
    B_p[\Gamma_L,\Gamma_R]
    &=
	\frac{1}{2}
	\!\!\!\!\!\!
	\tikzinclude{Keldysh_vertex_parametrization-channeldec-Bp}
	\!\!\!
	\\
    B_t[\Gamma_L,\Gamma_R]
    &=
	-\quad
	\tikzinclude{Keldysh_vertex_parametrization-channeldec-Bt}
    \end{align*}
    \caption{Diagrammatic representation of the bubble functions in Eq.~\eqref{eq:parametrization:bubbles}.}
    \label{fig:parametrization:fully_parametrized_bubbles}
\end{figure}

Susceptibilities are obtained from $G^{(4)}$, Eq.~\eqref{eq:Keldysh_formalism:Field_theory:4point_function}, by contracting pairs of external legs and subtracting the disconnected parts \cite{rohringer_new_2013,eliashberg_transport_1962}. For the spin-$\updown$ and spin-$\upup$ components, we get
\begin{subequations}
    \label{eq:susceptibilities_postprocessed}
\begin{align}
    \chi_{a,\sigma\sigma', \omega_a}^{12|1'2'} =& 
        \int_{\nu}
        \Pi_{a, \omega_a\nu}^{12|1'2'} 
        +  \!\!
        \int_{\nu}
        \int_{\nu'}
        \Pi_{a, \omega_a\nu}^{14|1'4'}
        \Gamma_{\sigma\sigma', \omega_a\nu\nu'}^{34|3'4'}
        \Pi_{a, \omega_a\nu'}^{32|3'2'}
        ,
        \\
    \chi_{p,\sigma\sigma', \omega_p}^{12|1'2'} =& \nonumber
        \int_{\nu}
        \Pi_{p, \omega_p\nu}^{12|1'2'} (1 - \delta_{\sigma,\sigma'})
        \\
        &+  \!\!
        \int_{\nu}
        \int_{\nu'}
        \Pi_{p, \omega_p\nu}^{12|3'4'}
        \Gamma_{\sigma\sigma', \omega_p\nu\nu'}^{34|3'4'}
        \Pi_{p, \omega_p\nu'}^{34|1'2'}
        ,
        \\
    \chi_{t,\sigma\sigma', \omega_t}^{12|1'2'} =& \nonumber
        -\int_{\nu}
        \Pi_{t, \omega_t\nu}^{12|1'2'} \delta_{\sigma,\sigma'}
        \\
        &+  \!\!
        \int_{\nu}
        \int_{\nu'}
        \Pi_{t, \omega_t\nu}^{12|3'4'}
        \Gamma_{\sigma\sigma', \omega_t\nu\nu'}^{34|3'4'}
        \Pi_{t, \omega_t\nu'}^{34|1'2'}
        .
\end{align}
\end{subequations}
From these functions, we obtain physical susceptibilities as $\chi_{\mathrm{d}/\mathrm{m}} = \chi_{t, \uparrow\uparrow}
\pm
\chi_{t, \uparrow\downarrow}$,
or after exploiting spin and crossing symmetry, Eqs.~\eqref{eq:vertex_symmetries_and_parametrization:crossing_symmetry} and \eqref{eq:su2spin_relation}, 
\begin{subequations}
\begin{align}
\chi_{\mathrm{d}}^{12|1'2'} & = 
2\chi_{t, \updown}^{12|1'2'}
-
\chi_{a, \updown}^{21|1'2'},
\\
\chi_{\mathrm{m}}^{12|1'2'} & = 
-
\chi_{a, \updown}^{21|1'2'}.
\end{align}
\end{subequations}
These functions have the Keldysh structure of 4p functions. To identify the retarded susceptibilities $\chi^R(\omega)
$ in terms of 2p functions [analogous to the propagator, Eq.~\eqref{eq:G_Keldysh_structure}], we use the bare three-leg Hedin vertex $\lambda_0^{(k_1k_2) k_3}$ \cite{Hedin_1965} where the Keldysh indices $k_1$, $k_2$ belong to $\chi^{12|1'2'}$ and $k_3$ to $\chi^R$.
In terms of contour indices, it reads $\lambda_0^{(c_1 c_2) c_3} = -c_1 \delta_{c_1=c_2=c_3}$; in Keldysh indices, the nonzero components are
\begin{align}
    \lambda_0^{(kk)2} = \frac{1}{\sqrt{2}}
    =
    \lambda_0^{(k\bar{k})1} .
\end{align}
Hence, two (un-)equal fermionic Keldysh indices translate to a ``2'' (``1'') for the bosonic line.
We thus identify
\begin{align}
    \chi_{r}^{R} = \chi_{r}^{2|1} = 2\chi_{r}^{11|12}, 
    \quad 
    r=a,p,t.
\end{align}

In the parquet formalism, it was shown that the susceptibilities $\chi_r$ ($r \!\in\! \{a,p,t\}$) are related to asymptotic functions via \cite{wentzell_high-frequency_2020}
\begin{subequations}
    \label{eq:relation_of_susceptibilities_to_K1r}
\begin{align}
    (K_{1a})_{1'2'|12} &= - (\Gamma_{0})_{1'4'|32} (\chi_{a})_{34|3'4'} (\Gamma_{0})_{3'2'|14},
    \\
    (K_{1p})_{1'2'|12} &= - (\Gamma_{0})_{1'2'|34} (\chi_{p})_{34|3'4'} (\Gamma_{0})_{3'4'|12},
    \\
    (K_{1t})_{1'2'|12} &= - (\Gamma_{0})_{4'2'|42} (\chi_{t})_{34|3'4'} (\Gamma_{0})_{1'3'|13}.
\end{align}
\end{subequations}
For the retarded spin-$\updown$-component, we have
\begin{align}
    \label{eq:susceptibility_flowing}
    K_{1r\,\updown}^{R } 
    &= 
    - U^2 
    \chi_{r\,\updown}^{R} .
\end{align}
Although one-loop fRG does not fulfill the BSEs \eqref{eq:parquet:BSEa}-\eqref{eq:parquet:BSEt}, Eq.~\eqref{eq:susceptibility_flowing} can still be used as an estimate for susceptibilities. In the present context, these are often called ``flowing'' susceptibilities, while Eq.~\eqref{eq:susceptibilities_postprocessed} defines the ``post-processed'' susceptibilities. 
The PA, fRG, and K1SF results for $\chi_m$ and $\chi_d$ shown in the main text were computed using Eqs.~\eqref{eq:relation_of_susceptibilities_to_K1r}.

\section{Channel-adapted SDE}\label{app:SDE}
In the parquet formalism, the frequency dependence of the self-energy $\Sigma(\nu)$ enters via the second term in the SDE~\eqref{eq:mfRG:mfRG:Parquet:SDE}.
In the following, we discuss three options for the numerical evaluation of this diagram.

\begin{figure}
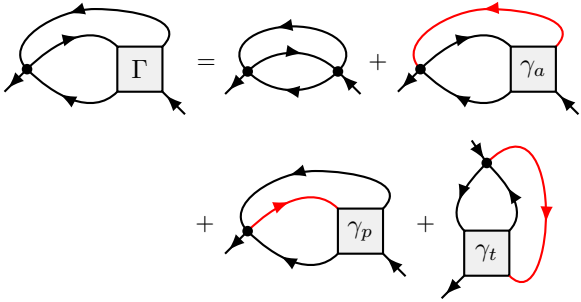

    \centering
    \begin{align*}
        \tikzinclude{mfRG-parquet-SDE_Gamma} &=
        \tikzinclude{mfRG-parquet-SDE_Gamma_0}
        +
        \tikzinclude{mfRG-parquet-SDE_gamma_a}
        \\
        &+
        \tikzinclude{mfRG-parquet-SDE_gamma_p}
        +
        \tikzinclude{mfRG-parquet-SDE_gamma_t}
    \end{align*}
    \caption{Rewriting of the SDE, where crossing symmetry was used for the $\gamma_t$ part. The red line indicates which propagator enters the loop $L$ in Eq.~\eqref{eq:SDE_versions_1}.
    }
    \label{fig:SDE_versions_as_diagrams}
\end{figure}

First,  using the parquet decomposition [Eq.~\eqref{eq:parquet_equation}], the second term of the SDE can be written in terms of bubbles $B_r$ and loop $L$ as (see Fig.~\ref{fig:SDE_versions_as_diagrams}) \cite{hille_quantitative_frg_2020,Eckhardt2020}
\begin{flalign}
\label{eq:SDE_versions_1}
\SigmaSDE{1} & = L(B_a(\Gamma_0, \Gamma_0), G) + \sum_r L(B_r(\Gamma_0,\gamma_r),\textcolor{red}G)
.
\hspace{-0.5cm} &
\end{flalign}
Here and below, a loop, $L$, acting on a $t$ bubble, $B_t$, contracts the two right legs, as opposed to the two top legs for all other vertex types (cf.~Fig.~\ref{fig:SDE_versions_as_diagrams}).

Second, the SDE in Eq.~\eqref{eq:mfRG:mfRG:Parquet:SDE}, without further manipulation, reads
\begin{align}
\label{eq:SDE_versions_2}
    \SigmaSDE{2} = L(B_r(\Gamma_0,\Gamma),G), \quad r\in\{a,p,t\},
\end{align}
where the channel $r$ can be freely chosen. 
Third, using $B_r(\Gamma_0, \gamma_r) = K_{1r} + K_{2'r}$ \cite{wentzell_high-frequency_2020}, the SDE equivalently reads
\begin{align}
\label{eq:SDE_versions_3}
    \SigmaSDE{3} = L(K_{1r} + K_{2'r},G).
\end{align}

Even though the above versions of the SDE are analytically equivalent, they vary in numerical accuracy and cost.
Evaluating $\SigmaSDE{3}$ is cheaper than the others since it skips the computation of bubbles $B_r$.
However, we found that Eq.~\eqref{eq:SDE_versions_1} is most accurate, since the $\gamma_r$ are inserted into bubbles $B_r$ of the same channel $r$.
Using the natural frequency parametrization for the reducible vertices $\gamma_r(\omega_r,\nu_r,\nu'_r)$, $\SigmaSDE{1}$ also has the benefit that one only needs to interpolate along the $\nu_r$-direction.

To illustrate this point, we consider a third-order contribution to the self-energy:
\begin{align} 
L(B_t(\Gamma_0,K_{1t}),\textcolor{red}G)
&= 
L(B_a(\Gamma_0,K_{1t}),\textcolor{red}G)
,
\\
\tikzinclude{Keldysh_formalism-Sigma_PT3_SDEv1_t}
&=
\tikzinclude{Keldysh_formalism-Sigma_PT3_SDEv2_t}
\, .
\end{align}
Inserting $K_{1t}$ into $B_{a}$ as done on the right results in diagrams that belong to the asymptotic class $K_{2'a}$.
However, on the left, $K_{1t}$ is inserted into $B_t$, resulting in diagrams belonging to $K_{1t}$. The latter can be treated with higher resolution and thus lead to better results for $\Sigma$, see Fig.~\ref{fig:SDE_versions}.
Note that the question how to best parametrize the SDE also arises in the context of the truncated-unity formalism for momentum-dependent models, where this choice was found to affect the quality of the results even more strongly due to the additional approximation from the truncation of the form-factor expansion \cite{hille_quantitative_frg_2020,Eckhardt2020}.

\begin{figure}
    \centering
    \includegraphics[scale=0.8]{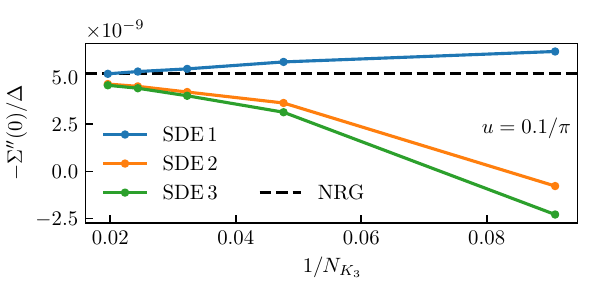}
    \caption{
    Imaginary part of the retarded self-energy at $\nu \!=\! 0$, computed with the parquet solver and different versions of the SDE, shown as function of $N_{K_3}$ ($u \!=\! 0.1/\pi$, $T/U \!=\! 0.01$). The dashed line is the NRG result. 
    For low $N_{K_3}$, SDE2 and SDE3 give the wrong sign. With increasing resolution, all results approach the correct value.}
    \label{fig:SDE_versions}
\end{figure}

\section{Equal-time correlators and Hartree self-energy}\label{app:equal-time}
Parts of the following discussion can be found in previous works, see Refs.~\onlinecite{jakobs_functional_2010, Weidinger2021, klockner_functional_2019}. We reiterate some of the points made there and extend on them to the context of this work.

The definitions of $G^{+|+}$  and $G^{-|-}$, Eqs.~\eqref{eq:G_jj'} and \eqref{eq:G_jj'_matrix}, are ambiguous at $t_1=t_2$ as $\Theta(t_1 \!-\! t_2 = 0)$ is not uniquely defined. If two operators $\psi, \psi^\dag$ are placed at the same point on the Keldysh contour, it is \textsl{a priori} not clear how to order them. The ambiguity is resolved by demanding that $\psi^\dag$ be put left of $\psi$ (``normal ordering''), which implies $G^{-|-}(t,t) = G^<(t,t) = G^{+|+}(t,t)$. Then, $G^< + G^> - G^{\mathcal{T}} - G^{\tilde{\mathcal{T}}} = 0$ does actually \textsl{not} hold, and care is due with Keldysh-rotated quantities. Since the point $t_1 \!=\! t_2$ is of zero measure in time integrals, which occur when computing diagrams in frequency space, this subtlety is irrelevant for most practical purposes. However, there is one important exception of equal-time nature, namely diagrams with loops that begin and end at the \textsl{same} bare vertex. With an instantaneous bare interaction, both incoming and outgoing legs have the same time argument, so that these diagrams involve the frequency-integrated (i.e., equal-time) propagator.

The equal-time propagator determines the Hartree self-energy of the AM (e.g., in PT2 calculations),
\begin{align}
	\Sigma_\mathrm{H}
	=-
    \tikzinclude{Keldysh_formalism-SigmaH}
	. 
\label{eq:PT_Hartree}
\end{align}
Recall that, for the sAM ($\epsilon_d \!=\! -U/2$), the Hartree term is constant, $\Sigma_\mathrm{H}=U/2$, and can be absorbed into the bare propagator $G_0^R\rightarrow G_\mathrm{H}^R$, see Eq.~\eqref{eq:Hartree-propagator}. Subsequently, $G_\mathrm{H}^R$ is used for all computations involving bare propagators. In analogy, in the aAM, the bare propagator is replaced by the Hartree propagator, too. However, here, $\Sigma_\mathrm{H}$ is not constant and must be computed self-consistently (using, e.g., a simple bracketing algorithm), as it enters both sides of Eq.~\eqref{eq:PT_Hartree}. 
Now, a naive computation of the retarded component of this diagram after the Keldysh rotation (and in the frequency domain) would yield
\begin{align}
	\Sigma_\mathrm{H}^R &= \Sigma_\mathrm{H}^{1|2} 
	=-
    \tikzinclude{Keldysh_formalism-SigmaH_K}
	=
	-
	\underbrace{
    \tikzinclude{Keldysh_formalism-SigmaH_K1}
	}_{G^{1|1} \,\overset{?}{=}\, 0}
	-
	\tikzinclude{Keldysh_formalism-SigmaH_K2} \nonumber \\
	&\overset{?}{=} \frac{U}{2} \int \frac{\md \nu'}{2\pi i} \ G^K (\nu')
    . \label{eq:Hartree_naive}
\end{align}
This is, however, incorrect since $G^{1|1}(t|t) \neq 0$ after Keldysh rotation.
The correct result can be found by staying in the contour basis, using that, at equal times, only $\Sigma_\mathrm{H}^{-|-}(t,t) = -\Sigma_\mathrm{H}^{+|+}(t,t)$ is non-zero. Keldysh rotation yields $\Sigma_\mathrm{H}^R(t,t)$ = $\Sigma_\mathrm{H}^{-|-}(t,t)$, for which one has
\begin{align}
	\Sigma_\mathrm{H}^R = \Sigma_\mathrm{H}^{-|-}
	=-
    \tikzinclude{Keldysh_formalism-SigmaH_C}
	= 
 U \int \frac{\md \nu'}{2\pi i} \ G^< (\nu').
\label{eq:Keldysh_formalism:Field_theory:1P:Hartree_term}
\end{align}
To compute Eq.~\eqref{eq:Keldysh_formalism:Field_theory:1P:Hartree_term} in thermal equilibrium, one can relate $G^<$ to $G^R$ using the inverse Keldysh rotation and the FDT [Eq.~\eqref{eq:Keldysh_formalism:Field_theory:1P:FDT_GK}]
\begin{align}
    G^< (\nu) 
    &= \tfrac{1}{2} [ - G^R(\nu) + G^A(\nu) + G^K(\nu) ] \nonumber \\
    &= -2i\, n_F(\nu) \mathrm{Im} G^R (\nu)
    ,
    \label{eq:Keldysh_formalism:Field_theory:1P:FDT_Glesser}
\end{align}
with the Fermi function $n_F(\nu) \!=\! 1/(1 \!+\! e^{\nu/T})$.
This discussion of $\Sigma_{\mathrm{H}}$ also applies to the PA via the first term of the SDE~\eqref{eq:mfRG:mfRG:Parquet:SDE}
(the second vanishes for $|\nu|\to\infty$).

In fRG, $\Sigma_{\mathrm{H}}$ is generally renormalized throughout the flow, according to Eq.~\eqref{eq:mfRG:fRG:Flow_equations:Selfenergy_flow} for $\dot{\Sigma}$. In the limit $|\nu|\to\infty$, relevant for extracting the Hartree contribution, only those diagrams survive for which the in- and outgoing lines are attached to the same bare vertex:
\begin{align}
    \dot{\Sigma}_{\mathrm{H}} \
    &=
    -
     \lim_{\nu\rightarrow\pm\infty}
    \tikzinclude{mfRG-1l-dSigma}
    =
    -
    \tikzinclude{SE_H-flow_hartree}
    - 
    \tikzinclude{SE_H-flow_K1t}
    -
    \tikzinclude{SE_H-flow_K2t}
    \,. \label{eq:hartree-flow}
\end{align}
In practice, the Hartree contribution $\dot{\Sigma}_{\mathrm{H}}$ is not computed separately but is part of the full self-energy flow. 
There, equal-time propagators are single-scale propagators, occurring in the following contributions,
\begin{align}
    \dot{\Sigma} \
    &= \
    -
    \tikzinclude{mfRG-1l-dSigma}
    \supset
    -
    \tikzinclude{SE_H-flow_hartree}
    - 
    \tikzinclude{SE_H-flow_K1t}
    -
    \tikzinclude{SE_H-flow_K2t_p}
    .
\end{align}
However, in the context of this work, it turns out that these specific equal-time loops \textsl{can} be computed from just the Keldysh-component of the single-scale propagator, as in the naive calculation Eq.~\eqref{eq:Hartree_naive}. The reason is that, in the hybridization flow, 
the retarded component of the single-scale propagator asymptotically scales as $\sim 1/\nu^2$ for $\nu\rightarrow \pm \infty$, see Eq.~\eqref{eq:single-scale-hyb}. Using the FDT in the forms of Eqs.~\eqref{eq:Keldysh_formalism:Field_theory:1P:FDT_Glesser} and \eqref{eq:Keldysh_formalism:Field_theory:1P:FDT_GK}, we can write 
\begin{align}
    S^K(\nu) &= 2i \left[1-2n_F(\nu)\right]\mathrm{Im}S^R(\nu) \nonumber \\
    &= 2i\, \mathrm{Im}S^R(\nu) + 2 S^{<}(\nu).
\end{align}
When computing $\int\!\mathrm{d}\nu\, S^K(\nu)$, one can apply Cauchy's theorem to the first term, using its asymptotic behavior (see above). Closing the integration contour by an infinite semicircle in the upper half plane, avoiding the pole in the lower half plane, gives zero. Hence, in the hybridization flow, we have $\int\! \mathrm{d} \nu S^K(\nu) = 2 \int\! \mathrm{d} \nu S^<(\nu)$, and the subtlety discussed previously is irrelevant. Note that this argument may not apply to other regulators, where $S$ has a different expression. 

\section{Diagrammatic definition of PT2}
\label{app:PT2}
Following the previous discussion, the Hartree term in PT2 is determined self-consistently. The resulting Hartree propagator $G_\mathrm{H}$ then fulfills the Dyson equation
\begin{align}
    \tikzinclude{PT2_GH}
    &=
    \tikzinclude{PT2_G0}
    +
    \tikzinclude{PT2_Dyson}
    \,.
\end{align}
In these and the following diagrams, the Hartree propagator $G_\mathrm{H}$ is represented by a black line, whereas the light gray line denotes the bare propagator $G_0$. The dynamical part of the self-energy is computed from the first non-trivial term of the SDE, using $G_\mathrm{H}$,
\begin{align}
    \tikzinclude{mfRG-parquet-SDE} \equiv
    \Sigma - \Sigma_\mathrm{H} &= - \frac{1}{2}\, \tikzinclude{PT2_SE}.
\end{align}
The vertex in PT2 is given by the three diagrams
\begin{align}
    \Gamma - \Gamma_0 = 
    \tikzinclude{PT2_gamma_a}
    + \frac{1}{2}\,
    \tikzinclude{PT2_gamma_p}
    -\,
    \tikzinclude{PT2_gamma_t}
    \, ,
\end{align}
again evaluated with $G_\mathrm{H}$ in the internal lines. Susceptibilities are then computed from this vertex via the standard formula; for $\chi_a$, e.g., (again using $G_\mathrm{H}$ throughout)
\begin{align}
    \chi_a &= 
    \tikzinclude{PT2_chi_m_bare}
    \,
    +\,
    \tikzinclude{PT2_chi_m_bare_SE_upper}
    \,
    +\,
    \tikzinclude{PT2_chi_m_bare_SE_lower} \nonumber\\
    &\quad \, +
    \,
    \tikzinclude{PT2_chi_m}
    \, .
\end{align}
To obtain exactly the second-order contribution to the susceptibility, one insertion of the dynamical part of the self-energy into each line of the bubble term is required, which gives rise to the second and third diagrams shown.

We checked that, in the sAM at sufficiently low temperatures,
our numerical PT2 solution matches the analytic $T \!=\! 0$ results of Ref.~\onlinecite{yamada_perturbation_1975}
(Eqs.~(3.14) and (3.6)--(3.8) therein)
\begin{subequations}
\begin{flalign}
Z & = 1 - \big( 3 - \tfrac{1}{4} \pi^2 \big) u^2
, & \\
-\Sigma''(\nu)/\Delta & = \tfrac{1}{2} u^2 (\nu^2 \!+\! \pi^2 T^2)/\Delta^2,
\quad |\nu|, T \!\ll\! \Delta
, & \hspace{-0.5cm} \\
\tilde{\chi}_{\mathrm{m}/\mathrm{d}} 
& = 
\tfrac{1}{2} \big[ 1 \pm u + \big( 3 - \tfrac{1}{4} \pi^2 \big) u^2 \big]
. &
\end{flalign}
\end{subequations}

\section{Implementation details}
\label{app:implementation}

Below, we describe our choices for the implementation of the parquet and fRG solver, the sampling of continuous functions, and the performance-critical quadrature routine. In the process, we also discuss the numerical accuracy of our results.

The evaluation of bubble diagrams, Eq.~\eqref{eq:parametrization:bubbles}, is a major bottleneck in our methods.
However, computations for different external arguments can be distributed efficiently over multiple threads and compute nodes.
It also proved beneficial to vectorize the sum over internal Keldysh indices by reordering and combining Keldysh indices $k_i$ to Keldysh multi-indices $(k_m,k_n)$
\begin{align}
    \label{eq:Implementation:Keldysh_multi_indices}
    \Gamma^{k_{1'},k_{2'},k_{1},k_{2}} \mapsto 
    \begin{cases}
    \Gamma^{(k_{1'},k_{2}),(k_{2'},k_{1})},   & \text{for $a$-channel},  \\
    \Gamma^{(k_{1'},k_{2'}),(k_{1},k_{2})},   & \text{for $p$-channel},  \\
    \Gamma^{(k_{2'},k_{2}),(k_{1'},k_{1})},   & \text{for $t$-channel},
    \end{cases}
\end{align}
turning the Keldysh sum into an ordinary matrix product (which is optimized in common linear algebra libraries).
This pre-processing step enables us to efficiently fetch matrix-valued integrands and to perform sums over Keldysh indices and spins in an optimized manner.
It requires all Keldysh components to be present in the data, and, therefore, all of them are included in our computations.
Consequently, FDTs could not be exploited to gain performance benefits as they merely relate different Keldysh components. 
    
For the integrals over internal frequencies in Eqs.~\eqref{eq:parametrization:bubbles} and \eqref{eq:parametrization:loop}, we implemented an adaptive quadrature algorithm which picks sampling points based on a local error estimate and tolerance ($\epsilon_{\text{rel}} \!=\! 10^{-5}$).
With various vertex components, the evaluation of a vertex at a certain frequency is rather expensive. Therefore, we choose a quadrature algorithm that reuses the previous function evaluations when it refines the quadrature value on a subinterval (4-point Gauss--Lobatto rule with 7-point Kronrod extension) \cite{press_numerical_2007}.
Due to fine structures in the integrands, we found a higher-order quadrature rule to be beneficial for the convergence of the routine.
To help the algorithm find the structure in the integrand, we subdivide the integration interval at the expected positions of structure in the vertices or the propagators.
Quadrature of the integrand's tails at high frequency is performed numerically by means of a suitable substitution of the integration variable \cite{galassi_gnu_2009}.
For matrix-valued integrands, we use the sup norm $\|\cdot\|_\infty$ to compute the error estimate for the quadrature.

\begin{figure}
    \centering
    \includegraphics[width=.5\textwidth]{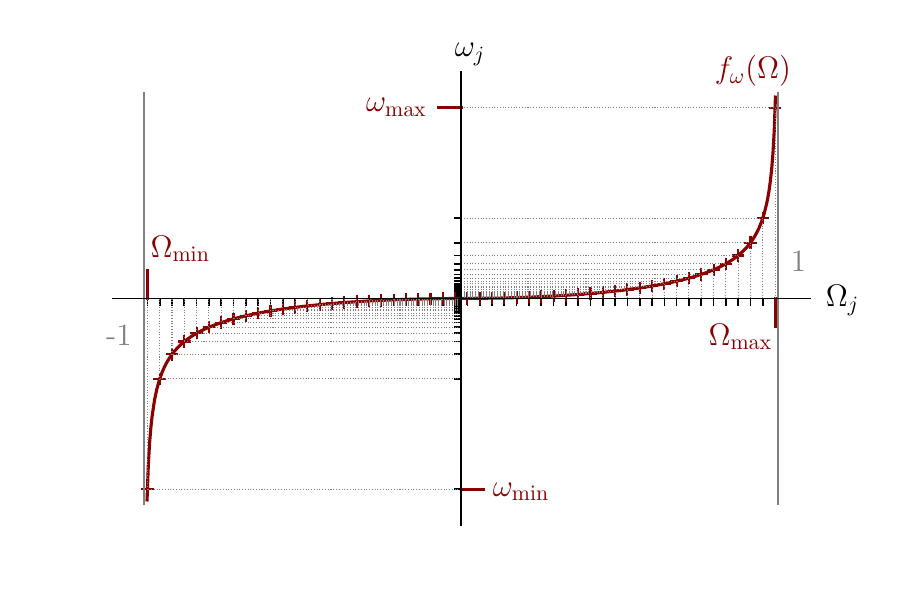}
    \caption{Non-linear frequency grid $\{\omega_j\}_{j=1}^N$ obtained via a transformation $f_A(\Omega)$, Eq.~\eqref{eq:implementation:grid_function}, from an auxiliary linear grid $\{\Omega_j\}_{j=1}^{N}$ of size $N$.}
    \label{fig:Implementation:Grid:Frequency_grid}
\end{figure}

Since Keldysh functions depend on continuous frequencies, a reliable and efficient representation is vital. We choose a non-uniform set of sampling points and obtain function values by (multi-)linear interpolation.
The overall behaviour of our functions is known:
The self-energy and the asymptotic functions $\Ktot{ir}$ can have sharp structures at smaller frequencies while, at large frequencies, they decay to a constant value with an approximate $\omega^{-k}$ with $k\in\mathbbm{N}$.
To capture this behaviour, we map an equidistant grid of an auxiliary variable $\Omega\in[-1,1]$ to a non-uniform one via the function
\begin{align}
\label{eq:implementation:grid_function}
    \omega = f_A(\Omega) = \frac{A \Omega |\Omega|}{\sqrt{1 - \Omega^2}}
\end{align}
with constant $A \!>\! 0$, see Figs.~\ref{fig:Implementation:Grid:Frequency_grid} and \ref{fig:Implementation:Grid:example_data}. 
The resulting sampling points are  dense around $\omega \!=\! 0$. At large frequencies, the function $f_A(\Omega)$ captures a $1/\omega^2$-decay effectively for $|\Omega| \!\lesssim\! 1$.
Furthermore, the structures in the AM scale approximately with the hybridization $\Delta$.
Therefore, we choose the frequency-grid parameter $A$ as multiples of $\Delta$ and ${\omega_{\text{max}} = 100 A}$.
With a fixed maximal frequency $\omega_{\text{max}}$, the variable $A$ determines the interval $[-\Omega_{\max}, \Omega_{\max}]$ used to construct the frequency grid via Eq.~\eqref{eq:implementation:grid_function}.
Our choices for $A$ are given in Tab.~\ref{tab:implementation:frequency_parameters}.

\begin{figure}
    \centering
    \includegraphics[scale=0.7]{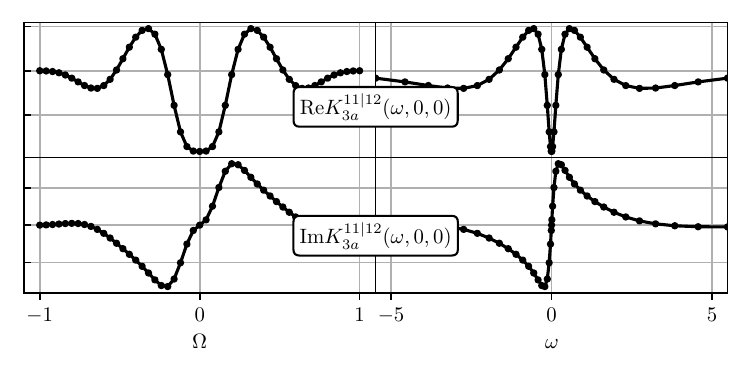}
    \caption{Illustration of the resolution of vertex data for a slice through  $\Re\Ktot{3a}^{11|12}$ and $\Im\Ktot{3a}^{11|12}$.
    The left panels show the data on the equidistant auxiliary grid, the right panels show the data on real frequencies. 
    Many sampling points are placed around the center where structures are peaked, while the tails are treated with very few points.
    Here, we also see an artifact due to our choice of the grid function \eqref{eq:implementation:grid_function}: since the grid function has a discontinuity at second order, we see a saddle point in the bottom left panel even though the function is linear there.
    The good resolution of the central peak in the real part comes at the cost of a saddle point in the imaginary part.
    }
    \label{fig:Implementation:Grid:example_data}
\end{figure}

\begin{table}[b]
\centering
\begin{tabular}{c|c|c|c|c|c|c}
& $\Sigma$ & $\Ktot1$  & $\Ktot{2,\omega}$  & $\Ktot{2,\nu}$  & $\Ktot{3,\omega}$  & $\Ktot{3, \nu}$ \\
\hline
$A/\Delta$    & 10        & 5         &   15              &   20             &10                     &   10
\end{tabular}
\caption{Frequency-grid parameter $A$ for Eq.~\eqref{eq:implementation:grid_function}.}
\label{tab:implementation:frequency_parameters}
\end{table}

It is also possible to adapt the frequency-grid parameter $A$ automatically.
Interpolating the vertex linearly, we can approximate the error by the maximal curvature in the  space of the linearly sampled auxiliary variable $\Omega$.
Hence, we can use the curvature as an error function to optimize the parameter $A$ in Eq.~\eqref{eq:implementation:grid_function}.
The direction-dependent curvature of a multi-variate function $f$ is encoded in the Hessian, $H_{ij} = \partial_i\partial_j f(\vec{x})$. We can efficiently compute a scalar measure for the curvature via the Frobenius norm of the Hessian, giving
\begin{align}
\|H\|_F^2 = \sum_{i,j} |H_{i,j}|^2 = \text{Tr} H^2 = \sum_i |\lambda_i|^2 ,
\end{align}
where $\lambda_i$ are the eigenvalues of $H$.
An approximation of the partial derivatives can be obtained with the finite differences method.
However, for the studied parameter regime of the AM, we found (using Brent's method \cite{brent_algorithms_2013} as the minimizer) that optimizing the grid parameters $A$ did not make a big difference compared to a simple rescaling according to Tab.~\ref{tab:implementation:frequency_parameters}.
To verify convergence in the number of sampling points, we compared the static susceptibilities between implementations in the KF and the MF and found agreement up to 1\textperthousand, see Fig.~\ref{fig:convergence_in_resolution}.

\begin{figure}
    \centering
    \includegraphics[width=\linewidth]{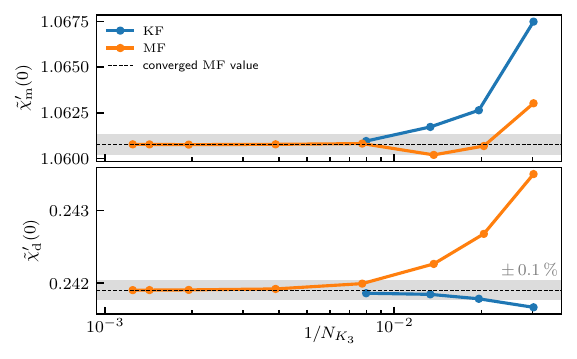}
    \caption{
    Convergence w.r.t.\ frequency resolution for the static susceptibilities as in Fig.~\ref{fig:stat_susc} from parquet solvers in the KF and the MF for $u \!=\! 0.75$ (a setting where $K_{2}$ and $K_{3}$ are relevant).
    The numbers of frequency points for $K_{1}$ and $K_{2}$ are chosen proportional to $N_{K_3}$.
    In the MF, we used $N_{K_3} \!=\! 33, 49, 73, 129, 257, 513, 701, 801$, in the KF $N_{K_3} \!=\! 33, 51, 75, 125$.
    The KF and MF results agree very well;
    the shaded region marks {1\textperthousand} deviation.
    }
    \label{fig:convergence_in_resolution}
\end{figure}

To solve the fRG equations \eqref{eq:mfRG:fRG:Flow_equations} we employ a Runge--Kutta solver with adaptive step size control (Cash--Carp). The step size is chosen according to an error estimate and tolerance (here: relative error $\epsilon_{\text{rel}} \!=\! 10^{-6}$). 
Furthermore, we reparametrize the flow parameter $\Lambda(t) = f_{A=5}(t)$ to provide a good first guess for the step sizes, using the same function $f_A(t)$ as for frequencies $\omega$, Eq.~\eqref{eq:implementation:grid_function}, with $A \!=\! 5$. It provides large steps for high $\Lambda$ and small steps for small $\Lambda$ for equidistant $t$.
As initial condition of $\Sigma^{\Lambda_i}$ and $\Gamma^{\Lambda_i}$ at large $\Lambda_i$, we use the converged parquet solution. As discussed in Sec.~\ref{sec:results}, the PA gives good results in the perturbative regime.

To solve the self-consistent parquet equations $f_{\text{PA}}$ in Eqs.~\eqref{eq:all_parquet_equations}, which constitute a fixed-point equation for the state $\Psi=(\Sigma, \Gamma)$, i.e., $\Psi=f_{\text{PA}}(\Psi)$,
we perform fixed-point iterations until the result meets a tolerance criterion, here ${\|\Psi - f_{\text{PA}}(\Psi)\|_\infty < 10^{-6} \|\Psi\|_\infty}$.
For intermediate to higher $u\gtrsim 1$, it proves beneficial to stabilize the algorithm with a partial update scheme, \ie 
\begin{align}
    \Psi &\leftarrow (1-m) \cdot \Psi + m \cdot f_{\text{PA}}(\Psi)
\end{align}
with mixing factor $0<m\leq1$ (here typically $m=0.5$). 
For faster convergence in the vicinity of the fixed point, we use Anderson acceleration \cite{anderson_iterative_1965,walker_anderson_2011}. 

\section{Numerical costs}\label{app:numerics}
The numerically most complex objects in all calculations are the $K_3$ components of the two-particle reducible vertices, as they depend on three continuous frequency arguments independently. The numerical cost of a parquet or fRG computation is therefore $\mathcal{O}(N_{K_3}^3)$, where $N_{K_3}$ is the number of grid points per frequency used for $K_3$. This applies to memory (as all this data has to be stored) and to computation time  (as BSEs or fRG flow equations are evaluated for all external arguments). We give in Tab.~\ref{tab:frequency_points} the number of frequency points used for each diagrammatic class. The self-energy was resolved on a grid with the same number of points as the $K_1$ class.

\begin{table}[t]
    \centering
    \begin{tabular}{c|c|c|c}
        & $N_{K_1}$ & $N_{K_2}$ & $N_{K_3}$  \\ \hline
        fRG & $401$ & $201$ & $101$ \\
        PA & $401$ & $201$ & $51$--$101$ \\
        PT2 & $801$ & $0$ & $0$ \\
        \SF & $401$ & $0$ & $0$
    \end{tabular}
    \caption{Number of frequency points for different diagrammatic classes and methods. We use the same number of points for $\Sigma$ as for $K_1$. In most PA computations, $N_{K_3}=51$, except for the largest values of $u$, which required $N_{K_3} \!=\! 101$ for converging the parquet solver.}
    \label{tab:frequency_points}
\end{table}

The numerical cost is further determined by the accuracy (or the convergence criteria) chosen for the iterative parquet solver or the Runge--Kutta solver in fRG flow (see App.~\ref{app:implementation}). Finally, the accuracy of the integrator also affects the numerical cost strongly (see again App.~\ref{app:implementation}). Our most costly computations were $150$ iterations of the parquet solver with $N_{K_3} = 101$ (required for convergence in the region $u\lesssim 1$). On the KCS cluster at the LRZ, equipped with chips of the type \textsl{Intel$^\circledR$ Xeon$^\circledR$ Gold 6130 CPU @ 2.10GHz} capable of hyper-threading, one such computation took about two days on $32$ nodes, running $32$ threads each.

\section{Convergence of $\tilde{\chi}_\mathrm{m}(0)$}\label{app:chi_m_convergence}
\begin{figure}
    \centering
    \includegraphics[width=\linewidth]{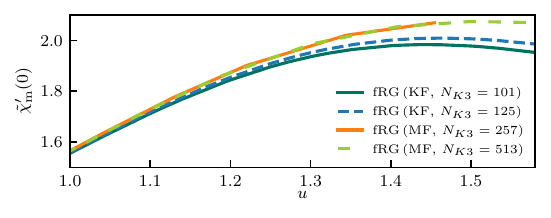}
    \caption{
    Static magnetic susceptibility of the sAM obtained with fRG. Compared to Fig.~\ref{fig:stat_susc}, there is an additional KF (MF) line with higher (lower) resolution. The MF result appears converged in $N_{K_3}$; the KF result is slightly improved by increasing $N_{K_3}$ from 101 to 125.}
    \label{fig:chi_m_convergence}
\end{figure}
Figure \ref{fig:chi_m_convergence} shows the static magnetic susceptibility of the sAM obtained with fRG, zooming into the regime $u\gtrsim 1$ (where deviations between MF and KF results become noticeable) and scrutinizing convergence w.r.t.\ frequency resolution. Compared to Fig.~\ref{fig:stat_susc}, there is an additional KF (MF) line with higher (lower) resolution, as determined by the number of frequency points used to resolve the $K_3$ class, $N_{K_3}$ (cf.\ Fig.~\ref{fig:convergence_in_resolution}).
The MF result appears converged in $N_{K_3}$, whereas the KF result is slightly improved by increasing $N_{K_3}$. The improvement is minor, however, and does not justify the additional numerical cost: The computation for $N_{K_3}=125$ consumed roughly 30,000 CPUh, while the computation for $N_{K_3}=101$ took only half that time. Nevertheless, one should keep in mind that these computations yield a full parameter sweep in $u$ and are thus more economical than individual PA computations. Further analysis, including line plots through all vertex components and asymptotic classes, is provided in the data set attached to this paper. This analysis shows that the resolution of fine structures in some Keldysh components of the $K_3$ class could still be improved using even higher values of $N_{K_3}$.

\bibliography{bibliography.bib}

\end{document}